\documentclass[11pt]{article}
\usepackage[T1]{fontenc}
\usepackage[utf8]{inputenc}
\usepackage{times}
\usepackage{latexsym}
\usepackage{marvosym}
\usepackage[dvipsnames,table,xcdraw]{xcolor}
\usepackage[preprint]{acl}

\usepackage{amsmath,amsfonts,bm}

\def\eqref#1{equation~\ref{#1}}

\def\1{\bm{1}}

\DeclareMathAlphabet{\mathsfit}{\encodingdefault}{\sfdefault}{m}{sl}
\SetMathAlphabet{\mathsfit}{bold}{\encodingdefault}{\sfdefault}{bx}{n}

\usepackage{url}

\usepackage[linesnumbered,ruled,lined]{algorithm2e}
\usepackage{booktabs}
\usepackage{multirow}
\usepackage{graphicx}
\usepackage{array}
\usepackage{makecell}
\usepackage{enumitem}
\usepackage{microtype}
\usepackage{inconsolata}
\usepackage{balance}
\usepackage{placeins}
\usepackage[most]{tcolorbox}
\usepackage{fancyvrb}
\usepackage{tabularx}
\usepackage{pgf}

\hypersetup{
    pdftitle={DeepRepoQA: Code Repository Question Answering with Deep Agent Exploration},
    pdfauthor={Weihan Peng, Yuling Shi, Yingwei Ma, Longfei Yun, Beijun Shen, and Xiaodong Gu}
}

\newcommand{\poschangenew}[1]{%
    \pgfmathsetmacro{\myval}{#1}%
    \pgfmathtruncatemacro{\percent}{min(100, 50 + \myval * 15)}%
    (\textcolor{ForestGreen!\percent}{+#1})%
}

\newcommand{\negchangenew}[1]{%
    \pgfmathsetmacro{\myval}{#1}%
    \pgfmathtruncatemacro{\percent}{min(100, 50 + \myval * 120)}%
    (\textcolor{red!\percent}{-#1})%
}

\newcommand{\ourmethod}[0]{\textsc{DeepRepoQA}\xspace}
\newcommand{\ourdata}[0]{\textsc{SWE-QA}\xspace}

\newcounter{prompt}

\begin{document}

\title{\ourmethod: Code Repository Question Answering with Deep Agent Exploration}

\author{
  Weihan Peng\textsuperscript{1} \quad Yuling Shi\textsuperscript{1} \quad
  Yingwei Ma\textsuperscript{2} \quad Longfei Yun\textsuperscript{3} \\
  \bfseries Beijun Shen\textsuperscript{1} \quad Xiaodong Gu\textsuperscript{1,\Letter} \\
  \normalfont \textsuperscript{1}Shanghai Jiao Tong University \\
  \textsuperscript{2}The Hong Kong University of Science and Technology \\
  \textsuperscript{3}University of California San Diego
}

\maketitle
\begingroup
\renewcommand{\thefootnote}{}
\footnotetext{\Letter\ Corresponding author}
\endgroup
\begin{abstract}
Answering developer questions about a software repository is a critical yet under-explored problem in software engineering.
While existing repository understanding methods have advanced the field, they predominantly rely on surface-level code retrieval and lack the ability for deep reasoning over multiple files, complex software architectures, and grounding answers in long-range code dependencies.
To address these limitations, we propose \ourmethod, a novel question answering (QA) framework for repository-level code understanding. \ourmethod builds on the agentic framework where LLM agents find answers through a systematic tree search over the repository structure. 
A Monte-Carlo Tree Search (MCTS) mechanism is employed to empower agents to dynamically search, navigate, and inspect code, enabling effective multi-hop reasoning over long-range code dependencies. 
Comprehensive experiments on the \ourdata benchmark demonstrate substantial performance gains over strong baselines, validating the effectiveness of systematic MCTS-guided exploration for multi-hop repository reasoning.
\end{abstract}

\section{Introduction}
\label{sec:intro}
\begin{figure*}[t]
    \centering
    \includegraphics[width=0.95\linewidth]{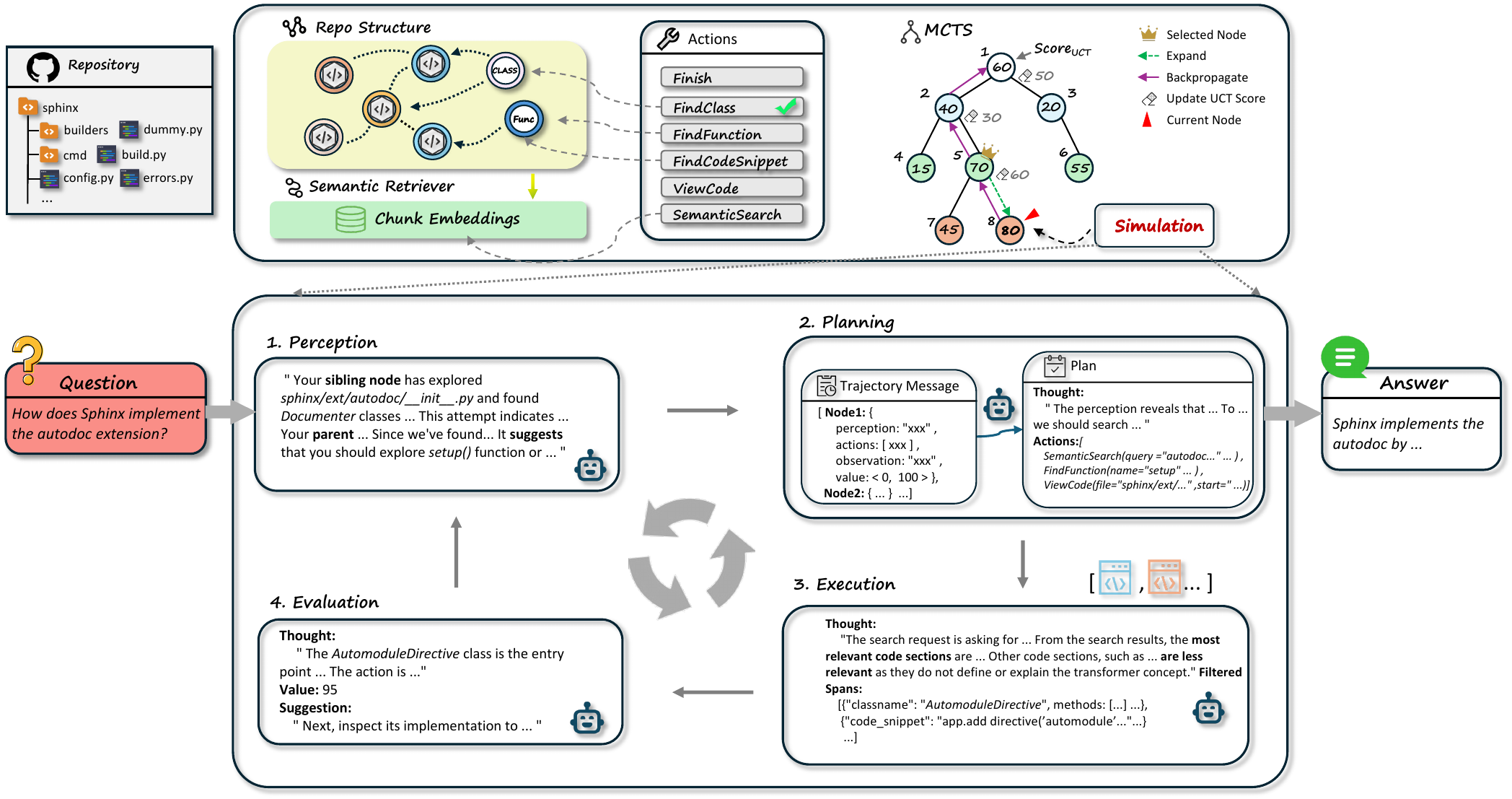}
    \caption{Overview of the \ourmethod framework.
    }
    \label{fig:approach}
\end{figure*}
Answering questions about a software repository is an underexplored yet critical problem~\cite{CodeQA,peng2026swe,chen2025coreqa}. To accomplish a function or resolve an issue, developers frequently need to navigate large, interconnected codebases, trace dependencies across multiple files, and synthesize architectural knowledge in order to answer complex questions~\cite{jimenez2024swe,zhang2023repocoder,ouyang2024repograph}. With the rapid advancement of LLMs~\cite{yao2023react,yang2024swe}, there is growing interest in repository-aware assistants that can provide end-to-end support for such tasks.

Despite rapid progress, existing repository understanding methods often operate at the function level~\cite{CoSQA,CodeQA,CS1QA,li2024procqa,CodeQueries} or rely on basic retrieval-augmented generation (RAG) that surfaces local fragments of code~\cite{zhang2023repocoder}. These approaches are ill-suited to deep repository reasoning, where answers require following call chains across files, combining dispersed evidence, and articulating architectural intent~\cite{ouyang2024repograph}. ReAct-style agents~\cite{ReAct}, such as OpenHands~\cite{wang2024openhands}, address some of these limitations by introducing single-path, reasoning-augmented interaction, allowing models to interleave action execution and reasoning for multi-step question answering. While they enable basic planning and verification, their single-path design still constrains exploration: important evidence may be missed if it lies outside the chosen trajectory, and reasoning depth is limited by sequential execution. Moreover, balancing exploration and exploitation remains challenging, which can result in retrieval bias or incomplete synthesis of cross-file dependencies.

To address these limitations, we propose \ourmethod\footnote{\url{https://github.com/peng-weihan/DeepRepoQA}}, a novel agentic question answering (QA) framework for repository-level deep understanding. 
Unlike flat retrieval pipelines~\cite{zhang2023repocoder}, \ourmethod casts QA as a deep search-and-verify process over the repository structure, enabling LLM agents to systematically explore repositories and synthesize evidence before producing answers. The framework defines a compact, semantically meaningful action space spanning semantic search, structural navigation, and targeted code inspection. 
Given a question, the agent initializes a search tree and iteratively (i) perceives the current state and reasoning trajectory, (ii) plans the next step, (iii) executes the action over an index-backed representation to gather verifiable evidence, and (iv) evaluates utility to steer future exploration. 
Collected evidence is curated and appended to a memory path from the root to the current node; this path structures prompts, stabilizes reasoning, and, once sufficient evidence is collected, enables synthesis of a final answer grounded in cited code spans. 
Compared to single-path ReAct-style agents~\cite{peng2026swe}, our approach introduces two key innovations: (1) an iterative search-and-verify process via MCTS for multi-hop repository reasoning, and (2) learned priors/values from LLM feedback that reduce search depth and drift. Together, these elements transform repository QA from ad-hoc retrieval to systematic search with consistent, evidence-based answers. 

We evaluate \ourmethod on the \ourdata benchmark~\cite{peng2026swe} across multiple LLM backbones, including Kimi K2, GLM-4.6, Qwen3-Coder-480B-A35B-Instruct and GPT-5.1. \ourmethod achieves consistent improvements over strong baselines. On all models, \ourmethod outperforms existing state-of-the-art ReAct-style agents, with the largest gain of 7.08 points over SWE-agent on Qwen3-Coder-480B-A35B-Instruct. \ourmethod also surpasses or matches leading commercial tools such as Cursor and Tongyi Lingma when using GPT-5.1 and Kimi K2.


Ablations highlight the critical importance of MCTS, while semantic search, Evaluation Agent, and Perception Agent further enhance stability and efficiency. Efficiency analysis demonstrates that \ourmethod achieves strong performance in both computational cost and answer quality. Furthermore, a case study shows that our MCTS-guided, verification-centric exploration mitigates retrieval bias by grounding answers in cross-file evidence.

Our contributions are as follows:

\begin{itemize}[leftmargin=20pt]
    \item We propose an agentic QA framework for repository understanding, enabling structured, verifiable multi-hop reasoning over long-range code dependencies.
    \item We introduce MCTS-guided exploration with LLM feedback and memory for efficient, repository-grounded search.
    \item We conduct a comprehensive evaluation on \ourdata, demonstrating consistent improvements over state-of-the-art baselines.
\end{itemize}
\section{Related Work}

Repository-level code understanding has become a central challenge for modern software development, such as automated code generation~\cite{shrivastava2023repofusion,zhang2024codeagent,shi2024code}, code translation~\cite{wang2024repotransbench,wang2025evoc2rust}, and issue resolution~\cite{jimenez2024swe,chen2025swe}. These approaches include retrieval-augmented context gathering~\cite{zhang2023repocoder}, agent-based repository navigation~\cite{ma2024understand,yang2024swe,li2025swe}, graph-based cross-file relation modeling~\cite{ouyang2024repograph}, and context-aware code completion~\cite{shrivastava2023repofusion}.
LongCodeZip compresses long code contexts by retaining task-relevant functions and code blocks~\cite{shi2025longcodezip}.
While highly effective for their intended purposes, they primarily focus on code synthesis and modification, rather than on the comprehensive, multi-hop reasoning required for deep question answering.

In code question answering, most existing work targets snippet or function-level understanding. Studies include neural QA systems for subroutine behaviors~\cite{BansalEWM21}, task-adaptive pre-training for code QA~\cite{yu2022code}, and transformer-based models for regulatory codes~\cite{xue2024question}. Recently, attention has shifted to repository-level QA, which addresses architectural questions requiring multi-file, multi-hop reasoning across entire codebases. While benchmarks and improvements have been proposed~\cite{chen2025coreqa,strich2024improving,peng2026swe}, the field remains nascent. SWE-Bench ProMax benchmarks agents on expert-curated, multilingual, large-scale code-refactoring tasks~\cite{shi2026swe}. Analyses indicate that directly applying LLMs and RAG to repository-level QA faces inherent limitations~\cite{Georgy2024}, highlighting the need for more structured reasoning approaches.


Comparatively, our \ourmethod overcomes these limitations by formulating repository QA as a planning problem. Using a multi-module Monte-Carlo Tree Search (MCTS), our framework enables a principled and systematic exploration of the vast and complex reasoning space within a codebase. This supports deliberate, multi-hop inference across entire repositories, surpassing simple retrieval or localized analysis. 

\section{Methodology}


We propose \ourmethod, a repository QA method that leverages deep agent exploration. 
\ourmethod adopts an agentic framework grounded in Monte-Carlo Tree Search (MCTS), where four specialized agents, responsible for perception, planning, execution, and evaluation, collaborate to enable deep exploration of the codebase. 
As illustrated in Figure~\ref{fig:approach}, \ourmethod answers repository questions through an iterative search-and-verify loop. Upon receiving a question, the system constructs a compact search tree representing potential next steps (e.g., search, navigate, inspect, or stop). 
Each step in this loop consists of four stages governed by the corresponding agents: a \emph{Perception Agent} analyzes the current state of exploration. A \emph{Planning Agent} then proposes promising subsequent actions based on the perceived context. An \emph{Execution Agent} performs these actions to gather concrete evidence from the code. Finally, an \emph{Evaluation Agent} assesses the results, assigns value scores, and propagates feedback to guide subsequent exploration. This cycle repeats until sufficient evidence has been accumulated, after which the system synthesizes a final answer grounded in explicit code citations. 
Through this structured, feedback-driven process, \ourmethod transforms repository QA from a linear retrieval task into a multi-path reasoning process, thereby enabling effective handling of complex, multi-hop questions.
Detailed descriptions of each agent are provided in the following sections.

\subsection{Repository Parsing and Representation}
DeepRepoQA's ability to comprehend and reason over large codebases relies on a systematic \emph{repository parsing} and \emph{representation} pipeline. The goal of this pipeline is to transform the raw repository into structured information that can be efficiently queried and analyzed by the model.

\subsubsection{Repository Parsing} 
\label{sec:parsing}
The parsing process involves traversing the files and directory structure of the repository to identify and extract relevant source code elements. Specifically, DeepRepoQA analyzes files to detect classes, functions, and code snippets, capturing their essential syntactic and semantic information. This process leverages language-specific parsers and heuristics based on common project layouts (e.g., \texttt{src}, \texttt{tests}), and employs Tree-sitter\footnote{https://tree-sitter.github.io/tree-sitter/} to obtain lightweight, language-agnostic Abstract Syntax Trees that support precise structural extraction. Parsing also includes resolving cross-file relationships, such as function calls, class inheritance, and module dependencies, enabling the agent to build a comprehensive view of the repository's overall architecture.
 
\begin{figure}[t]
    \centering   
    \includegraphics[width=0.85\linewidth]{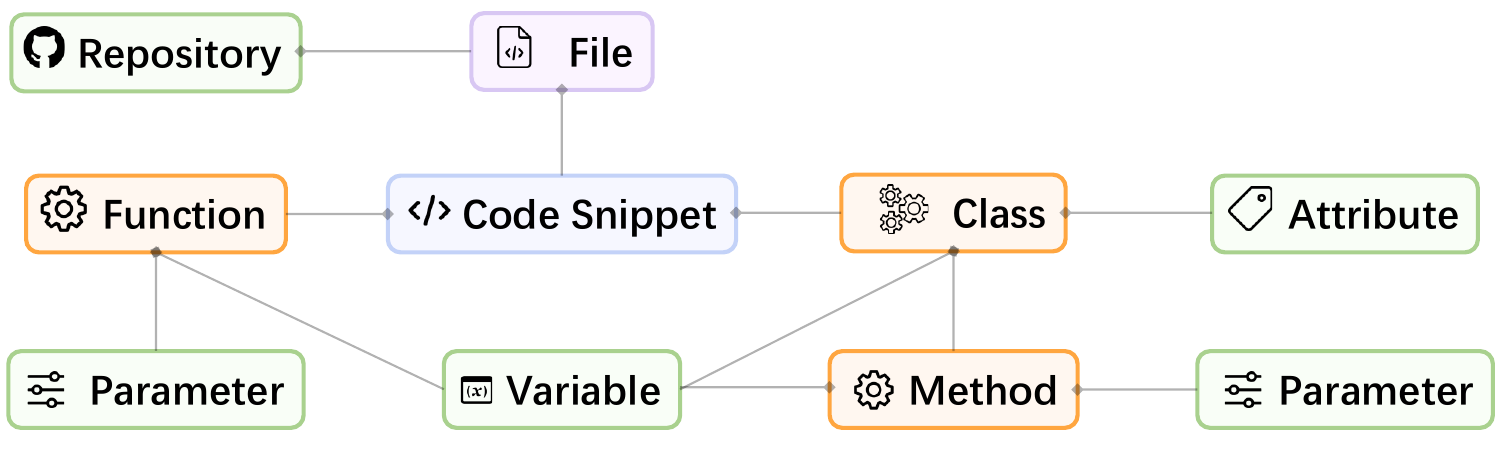}
    \caption{Core elements and their relations.}
    \label{fig:parse-overview}
\end{figure}
 
\subsubsection{Repository Representation} 
Once parsing is complete, the extracted information is transformed into structured representations suitable for downstream reasoning. DeepRepoQA employs a multi-level representation strategy:
\begin{itemize}[leftmargin=0.5cm]
    \item \textbf{Index-based Lookup:} The repository directory structure and AST trees are modeled to help the agent understand the positions and contexts of classes and functions within modules. By directly parsing ASTs, the agent can precisely locate classes, functions, and code snippets without relying on fuzzy or semantic search, enabling structure-aware retrieval via \texttt{FindClass}, \texttt{FindFunction}, and \texttt{FindCodeSnippet}.
    \item \textbf{Semantic Retriever:} Code elements requiring conceptual-level retrieval (e.g., arbitrary code snippets) are vectorized to support semantic search and reasoning. Indexed nodes are embedded and queried using a RAG-style Semantic Retriever, corresponding to the \texttt{SemanticSearch} operation in the system.
\end{itemize}
By combining these representations, DeepRepoQA achieves both \emph{high-level structural understanding} and \emph{fine-grained semantic comprehension}, enabling effective repository analysis and intelligent question-answering.

\subsection{Action Space}
\label{sec:ActionSpace}

The DeepRepoQA agent operates with a discrete action space consisting of six core capabilities. These actions equip the agent with essential tools for repository comprehension, targeted code retrieval, and answer synthesis.

The action space is designed to be both comprehensive and concise, providing a powerful set of operations for repository exploration. Actions can be categorized into name-based retrieval (\texttt{FindClass}, \texttt{FindFunction}, and \texttt{FindCodeSnippet}), which enables precise lookups when specific identifiers are known, and semantic retrieval (\texttt{SemanticSearch}), which supports conceptual queries based on natural language. The \texttt{ViewCode} action allows the agent to inspect specific code regions returned by previous searches, while the \texttt{Finish} action signals the termination of the reasoning process. This structured action space guides the agent's reasoning and ensures that its interactions with the codebase are grounded, efficient, and interpretable. These actions and their corresponding arguments are summarized in Table~\ref{tab:action_options}.

\begin{table}[t]
\centering
\caption{Action Options in \ourmethod}
\label{tab:action_options}
\resizebox{\columnwidth}{!}{
\begin{tabular}{>{\raggedright\arraybackslash}p{3cm} >{\raggedright\arraybackslash}p{5cm} c}
\toprule
\textbf{Action} & \textbf{Arguments} & \textbf{Embedding?} \\
\midrule
\textbf{FindClass} & class-name, file-pattern & No \\
\textbf{FindFunction} & function-name, class-name, file-pattern & No \\
\textbf{FindCodeSnippet} & code-snippet, file-pattern & No \\
\textbf{ViewCode} & code-span-list, file-pattern & No \\
\textbf{SemanticSearch} & query, category, file-pattern & Yes \\
\textbf{Finish} & answer, finish-reason & No \\
\bottomrule
\end{tabular}
}
\end{table}

We now describe each action in detail as follows:

\begin{itemize}[leftmargin=0.5cm]
    \item \textbf{FindClass}: enables the agent to locate class definitions within the repository. This action supports high-level structural reasoning by identifying key abstractions, inheritance relationships, and encapsulated functionalities. 
    \item \textbf{FindFunction}: allows the agent to identify specific function or method definitions. This facilitates fine-grained understanding of code behavior, enabling the agent to trace logic, inputs, and outputs within targeted modules.
    \item \textbf{FindCodeSnippet}: provides the agent with the ability to extract relevant code fragments based on contextual needs. This action is essential for gathering actionable examples, verifying implementation details, or focusing on critical sections within large files.
    \item \textbf{SemanticSearch}: empowers the agent to perform advanced search over repository content using semantic embeddings rather than mere keyword matching. Leveraging a code embedding model, this action allows the agent to retrieve conceptually related code or documentation.
    \item \textbf{ViewCode}: grants the agent the ability to inspect the full contents of specific files. This low-level inspection is useful for understanding implementation details, configuration, and documentation. Standard command-line utilities such as cat or grep may be executed to facilitate precise content access.
    \item \textbf{Finish}: marks the completion of the answer task, indicating that the agent has synthesized a definitive answer from the collected information or reaches a predefined maximum number of iterations.
\end{itemize}

\subsection{Agents}
\subsubsection{Perception Agent}
The \emph{Perception Agent} analyzes the \emph{current search state} before any new action is proposed, serving as the sensory input for the MCTS framework. It synthesizes the trajectory from the root to the current node, along with information from sibling branches, to produce a concise situational report. This report serves multiple purposes: it detects redundancy to avoid re-exploring dead ends, highlights unvisited but promising code regions, surfaces conflicts or gaps in the collected evidence, and extracts structured cues (e.g., candidate symbols, files, or API calls) for further investigation. For a question like, ``How does Sphinx implement the autodoc extension?'', if one branch has explored \texttt{sphinx/ext/autodoc/\_\_init\_\_.py} and found \texttt{Documenter} classes, the Perception Agent may notice that the \texttt{setup()} function, the standard entry point for Sphinx extensions, has not been analyzed. It will therefore flag the \texttt{setup()} function as a target to guide the next planning phase toward discovering how the extension integrates with Sphinx's core. By delivering a compact, noise-reduced state assessment to downstream agents, it ensures that subsequent decisions are based on a holistic view of the search progress and blind spots. The prompt is provided in Appendix~\ref{sec:prompts}.

\subsubsection{Planning Agent}
The \emph{Planning Agent} is the core decision-maker, responsible for turning the situational report from the Perception Agent into a concrete action plan. When a leaf node is selected for expansion, this agent generates a set of candidate actions to create new child nodes, effectively growing the search frontier. It inspects the curated state summary to propose diverse and strategic next steps, while explicitly avoiding redundant trials already attempted in sibling branches. Building on the same example in which perception report flagged the \texttt{setup()} function as a key target, the Planning Agent might generate three candidate actions for expansion: 1) a \texttt{FindFunction} call for \texttt{setup} within the \texttt{sphinx/ext/autodoc/} directory, 2) a \texttt{SemanticSearch} for ``autodoc directive registration'' to find where the \texttt{automodule} directive is defined, and 3) a \texttt{ViewCode} action on \texttt{sphinx/ext/autodoc/\_\_init\_\_.py} to manually locate the \texttt{setup} function. Each proposed action becomes a new child node, representing a distinct reasoning path conditioned on both long-term goals and expected immediate information gain. The prompt is provided in Appendix~\ref{sec:prompts}.

\subsubsection{Execution Agent}
The \emph{Execution Agent} leverages the repository representations and action space introduced in Sections~\ref{sec:parsing} and~\ref{sec:ActionSpace} to carry out the planner-selected actions over the unified structural--semantic repository index. It produces raw outputs that include results from structure-aware lookup, semantic retrieval, and file-level inspection. Fuzzy name matching is applied to mitigate minor discrepancies in model-generated queries. Immediately after execution, the evidence-curation stage filters and normalizes the raw outputs: it ranks evidence by task relevance, de-duplicates overlapping spans, collapses boilerplate, and extracts focused quotations with exact file and line ranges. Low-value or noisy snippets are discarded, and salient context is stitched into a compact, citation-ready bundle. This curated evidence set becomes the observation for the current node and the context for subsequent reasoning. The detailed argument specifications and descriptions for each action are presented in Section~\ref{sec:ActionSpace}; the evidence-identification prompt is provided in Appendix~\ref{sec:prompts}.

\subsubsection{Evaluation Agent}
The \emph{Evaluation Agent} assesses the outcome of an executed action, providing critical feedback for the MCTS learning loop. After the Execution Agent returns a curated piece of evidence, this agent scores the action-observation pair by estimating a scalar value representing its utility in answering the original question. It also generates targeted, qualitative suggestions for subsequent steps, such as ``verify a dependency path'' or ``inspect call sites.'' For instance, if the \texttt{FindFunction} action for \texttt{setup} returns a code snippet containing the call \texttt{app.add\_directive('automodule', AutomoduleDirective)}, the Evaluation Agent would assign a high value (e.g., 95) to this node, as it directly reveals the link between the directive name and its implementing class. Its feedback might suggest, ``The \texttt{AutomoduleDirective} class is the entry point for the directive. Next, inspect its implementation to understand how it processes the directive's content.'' This value is then back-propagated up the search tree, updating the visit counts and value estimates of all nodes on the path from the current node to the root. This feedback loop allows the agent to continuously refine its search policy, prioritizing more promising reasoning paths. The prompt is provided in Appendix~\ref{sec:prompts}.

\subsection{Monte-Carlo Tree Search Process}
The four agents collaborate within the MCTS framework to answer a user's question. Starting from the question, \ourmethod builds a small search tree and repeatedly runs the four-agent simulation: the \emph{Perception Agent} forms a situational report, the \emph{Planning Agent} proposes an action, the \emph{Execution Agent} performs it with internal \emph{Filtering} to curate evidence, and the \emph{Evaluation Agent} scores utility and guidance. MCTS selects and expands promising branches until \texttt{Finish} is chosen, after which we synthesize a concise, citation-backed answer. The full pseudocode is provided in Algorithm~\ref{algo:approach}. The MCTS engine guides repository exploration through four phases---selection, expansion, simulation, and back-propagation---designed to systematically navigate the complex search space of repository-level reasoning. Appendix~\ref{app:mcts-workflow} provides a step-by-step illustration (Figure~\ref{fig:mcts-workflow}).

\begin{algorithm}[t]
    \caption{\ourmethod Algorithm}
    \label{algo:approach}
    \small
    \KwIn{User question $Q$, Repository context $R$, Max iterations $N$}
    \KwOut{Final answer $A$}
    search\_tree $\leftarrow$ MCTSTree($Q$, $R$)\;
    iteration\_count $\leftarrow 0$\;
    max\_iterations $\leftarrow N$\;
    \While{iteration\_count $<$ max\_iterations}{
        iteration\_count $\leftarrow$ iteration\_count + 1\;
        node $\leftarrow$ Select(search\_tree.root)\tcp*{Select promising node with highest UCT score, except those that have 3+ children}
        new\_node $\leftarrow$ Expand(node)\;
        situational\_report $\leftarrow$ Perceive(new\_node, new\_node.parent, new\_node.siblings)\;
        new\_node.action $\leftarrow$ PlanNextAction($Q$, situational\_report)\;
        raw\_result $\leftarrow$ ExecuteAction(new\_node.action)\;
        new\_node.observation $\leftarrow$ FilterEvidence($Q$, raw\_result)\;
        value $\leftarrow$ Evaluate(new\_node)\;
        Backpropagate(new\_node, value)\;
        \If{new\_node.action.type = \texttt{"Finish"}}{
            \textbf{break}\;
        }
    }
    finished\_nodes $\leftarrow$ GetFinishedNodes(search\_tree)\;
    $T \leftarrow$ GetBestTrajectory(finished\_nodes)\;
    $A \leftarrow$ ExtractAnswer($T$)\;
    \Return{$A$}
\end{algorithm}

\subsubsection{Selection}

Starting from the root node, child nodes are selected recursively according to the Upper Confidence Bound (UCT) policy until reaching a leaf node. The selection strategy balances exploration and exploitation using the UCT formula:
\begin{equation}
    \mathrm{UCT}(s, a) = Q(s,a) + c \cdot \sqrt{\frac{\ln N(s)}{N(s,a)}}
\end{equation}
where $Q(s,a)$ is the average value of taking action $a$ at state $s$, $N(s)$ is the visit count of $s$, $N(s,a)$ is the visit count of $(s,a)$, and $c$ is the exploration coefficient. In practice, nodes with higher UCT scores are prioritized for expansion. 
For example, as illustrated in Figure~\ref{fig:mcts-workflow}, node5, having the highest UCT score and being eligible for expansion, is selected.

\subsubsection{Expansion}

When selection reaches a node that is not fully expanded, a new child node is added. Before proposing candidates, the \emph{Perception Agent} produces a concise situational report from the direct trajectory and sibling attempts, highlighting redundancy, gaps, and promising regions. This lateral awareness lets planning avoid repeated trials without storing full sibling transcripts, since the perception report summarizes the relevant horizontal context. In practice, this expansion introduces a new branch for further exploration; for example, as illustrated in Figure~\ref{fig:mcts-workflow}, node5 is expanded to generate node8.

\subsubsection{Simulation}

Unlike traditional MCTS rollouts, our simulation is a single-step agent loop: \emph{Perception} summarizes state and siblings, \emph{Planning} proposes the next action, \emph{Execution} runs it and performs internal \emph{Filtering} to curate evidence, and \emph{Evaluation} assigns a scalar value. The evaluation output {$V(s,a)$} directly sets $Q(s,a) \leftarrow {V(s,a)}$, allowing effective assessment without multi-step rollouts and reducing compute while preserving decision quality. In practice, as illustrated in Figure~\ref{fig:mcts-workflow}, this single-step simulation is executed by the four agents sequentially, and the resulting evaluation value is used to update the corresponding state-action pair. For example, after the simulation, node8 receives a value of 80, which directly updates its $Q$ value.

\subsubsection{Back-propagation}

The simulation value is back-propagated to update $Q$ and visit counts along the path to the root. High-quality nodes are revisited more often, while low-value branches are implicitly pruned as their selection probability drops, focusing exploration on promising regions over time. The value obtained from the simulation is propagated back up the tree from the new node to the root, updating the visit counts and Q-values of all nodes along the path. This ensures that more promising branches are explored more frequently in subsequent iterations. For example, as illustrated in Figure~\ref{fig:mcts-workflow}, the value of node5 increases from 60 to 70, and the value of node2 increases from 30 to 40 during back-propagation. The process continues iteratively until termination.

\subsubsection{Answer Synthesis and Termination}

When the \texttt{Finish} action is selected, the system composes the final answer by summarizing the reasoning trace from the best trajectory and citing decisive evidence with file and line spans for traceability. We require at least one supporting code span to ensure grounding. The process concludes by validating that the answer addresses the question type (``what,'' ``where,'' ``how,'' or ``why'').

This iterative process allows DeepRepoQA to dynamically focus its search on the most promising reasoning paths, leading to more effective and efficient question answering.

\begin{table*}[t]
    \centering
    \caption{Overall Results on \ourdata 
    }
    \resizebox{0.92\linewidth}{!}{
    \begin{tabular}{l@{\hspace{3mm}}c@{\hspace{3mm}}c@{\hspace{3mm}}c@{\hspace{3mm}}c@{\hspace{3mm}}c@{\hspace{3mm}}c}
        \toprule
        \multirow{2}[2]{*}{\textbf{Methods}} & \multicolumn{5}{c}{\textbf{Evaluation Metrics}} & \multirow{2}[2]{*}{\textbf{Overall}} \\
        \cmidrule{2-6} 
        & \textbf{Correctness} & \textbf{Completeness} & \textbf{Relevance} & \textbf{Clarity} & \textbf{Reasoning} & \\ 
        \midrule
        \rowcolor{gray!20}\multicolumn{7}{l}{\textit{Commercial Tools}} \\
        \midrule
        \hspace{2mm}Tongyi Lingma 
        & 11.91 & 10.88 & 16.66 & 15.89 & 13.78 & 69.12 \\
        
        \hspace{2mm}Cursor
        & 12.31 & 11.34 & 17.32 & 16.07 & 13.67 & 70.71 \\
        
        \midrule
        \rowcolor{gray!20}\multicolumn{7}{l}{\textit{GLM-4.6}} \\
        \midrule
        \hspace{2mm}Direct Prompting 
        & 7.84 & 5.30 & 15.34 & 15.94 & 7.23 & 51.65 \\
        
        \hspace{2mm}Sliding Window RAG 
        & \;8.87 \poschangenew{1.03} 
        & \;7.11 \poschangenew{1.81} 
        & 14.06 \negchangenew{1.28} 
        & 15.95 \poschangenew{0.01} 
        & 10.01 \poschangenew{2.78} 
        & 56.00 \poschangenew{ 4.35} \\
        
        \hspace{2mm}Function Chunking RAG 
        & \;9.24 \poschangenew{1.40} 
        & \;7.45 \poschangenew{2.15} 
        & 14.02 \negchangenew{1.32} 
        & 15.97 \poschangenew{0.03} 
        & 10.22 \poschangenew{2.99} 
        & 56.90 \poschangenew{ 5.25} \\
        
        \hspace{2mm}SWE-agent 
        & 10.98 \poschangenew{3.14} 
        & 10.93 \poschangenew{5.63} 
        & 12.53 \negchangenew{2.81} 
        & 14.15 \negchangenew{1.79} 
        & 12.47 \poschangenew{5.24} 
        & 61.06 \poschangenew{ 9.41} \\
        
        \hspace{2mm}OpenHands 
        & 10.99 \poschangenew{3.15} 
        & 10.51 \poschangenew{5.21} 
        & 13.92 \negchangenew{1.42} 
        & \textbf{16.19} \poschangenew{0.25} 
        & \textbf{13.41} \poschangenew{6.18} 
        & 65.02 \poschangenew{13.37} \\
        
        \hspace{2mm}\ourmethod 
        & \textbf{11.11} \poschangenew{3.27} 
        & \textbf{11.30} \poschangenew{6.00} 
        & 13.43 \negchangenew{1.91} 
        & \textbf{16.65} \poschangenew{0.71} 
        & \textbf{13.41} \poschangenew{6.18} 
        & \textbf{65.90} \poschangenew{14.25} \\
                     
\midrule
\rowcolor{gray!20}\multicolumn{7}{l}{\textit{Kimi K2}} \\
\midrule
\hspace{2mm}Direct Prompting 
& 6.88 & 4.53 & \textbf{15.66} & 15.44 & 7.94 & 50.45 \\

\hspace{2mm}Sliding Window RAG 
& \;8.89 \poschangenew{2.01} 
& \;7.16 \poschangenew{2.63} 
& 15.07 \negchangenew{0.59} 
& \textbf{16.61} \poschangenew{1.17} 
& 10.22 \poschangenew{2.28} 
& 57.95 \poschangenew{ 7.50} \\

\hspace{2mm}Function Chunking RAG 
& \;9.04 \poschangenew{2.16} 
& \;7.50 \poschangenew{2.97} 
& 14.49 \negchangenew{1.17} 
& 16.51 \poschangenew{1.07} 
& 10.51 \poschangenew{2.57} 
& 58.05 \poschangenew{ 7.60} \\

\hspace{2mm}SWE-agent 
& 10.06 \poschangenew{3.18} 
& \;9.86 \poschangenew{5.33} 
& 14.56 \negchangenew{1.10} 
& 16.03 \poschangenew{0.59} 
& 11.85 \poschangenew{3.91} 
& 62.36 \poschangenew{11.91} \\

\hspace{2mm}OpenHands 
& 10.99 \poschangenew{4.11} 
& 10.35 \poschangenew{5.82} 
& 14.71 \negchangenew{0.95} 
& 15.68 \poschangenew{0.24} 
& 13.57 \poschangenew{5.63} 
& 65.30 \poschangenew{14.85} \\

\hspace{2mm}\ourmethod 
& \textbf{11.92} \poschangenew{5.04} 
& \textbf{11.08} \poschangenew{6.55} 
& 15.12 \negchangenew{0.54} 
& 16.31 \poschangenew{0.87} 
& \textbf{14.28} \poschangenew{6.34} 
& \textbf{68.71} \poschangenew{18.26} \\
           
        \midrule
        \rowcolor{gray!20}\multicolumn{7}{l}{\textit{Qwen3-Coder-480B-A35B-Instruct}} \\
        \midrule
        \hspace{2mm}Direct Prompting 
        & 7.71 & 5.85 & 14.85 & 15.00 & 7.92 & 51.33 \\
        
        \hspace{2mm}Sliding Window RAG 
        & 10.04 \poschangenew{2.33} 
        & \;8.22 \poschangenew{2.37} 
        & 15.06 \poschangenew{0.21} 
        & 15.83 \poschangenew{0.83} 
        & 10.98 \poschangenew{3.06} 
        & 60.13 \poschangenew{ 8.80} \\
        
        \hspace{2mm}Function Chunking RAG 
        & 10.01 \poschangenew{2.30} 
        & \;8.17 \poschangenew{2.32} 
        & \textbf{15.54} \poschangenew{0.69} 
        & 15.91 \poschangenew{0.91} 
        & 11.50 \poschangenew{3.58} 
        & 61.13 \poschangenew{ 9.80} \\
        
        \hspace{2mm}SWE-agent 
        & \;8.98 \poschangenew{1.27} 
        & \;8.49 \poschangenew{2.64} 
        & 13.26 \negchangenew{1.59} 
        & 15.60 \poschangenew{0.60} 
        & 11.02 \poschangenew{3.10} 
        & 57.35 \poschangenew{ 6.02} \\
        
        \hspace{2mm}OpenHands 
        & 10.30 \poschangenew{2.59} 
        & \;9.57 \poschangenew{3.72} 
        & 13.86 \negchangenew{0.68} 
        & 15.64 \poschangenew{0.64} 
        & 12.96 \poschangenew{5.04} 
        & 62.33 \poschangenew{11.00} \\
        
        \hspace{2mm}\ourmethod 
        & \textbf{10.97} \poschangenew{3.26} 
        & \textbf{10.17} \poschangenew{4.32} 
        & 13.82 \negchangenew{1.72} 
        & \textbf{16.19} \poschangenew{1.19} 
        & \textbf{13.28} \poschangenew{5.36} 
        & \textbf{64.43} \poschangenew{13.10} \\
                   
       \midrule
        \rowcolor{gray!20}\multicolumn{7}{l}{\textit{GPT-5.1}} \\
        \midrule
        \hspace{2mm}Direct Prompting 
        & 7.78 & 5.83 & 14.90 & 16.33 & 7.15 & 51.99 \\
        
        \hspace{2mm}Sliding Window RAG 
        & \;8.88 \poschangenew{1.10} 
        & \;6.24 \poschangenew{0.41} 
        & \textbf{16.89} \poschangenew{1.99} 
        & 16.03 \negchangenew{0.30} 
        & 10.15 \poschangenew{3.00} 
        & 58.19 \poschangenew{ 6.20} \\
        
        \hspace{2mm}Function Chunking RAG 
        & 10.15 \poschangenew{2.37} 
        & \;8.55 \poschangenew{2.72} 
        & 15.57 \poschangenew{0.67} 
        & \textbf{16.82} \poschangenew{0.49} 
        & 11.72 \poschangenew{4.57} 
        & 62.81 \poschangenew{10.82} \\
        
        \hspace{2mm}SWE-agent 
        & 12.12 \poschangenew{4.34} 
        & 10.61 \poschangenew{4.78} 
        & 15.06 \poschangenew{0.16} 
        & 14.55 \negchangenew{1.78} 
        & 13.05 \poschangenew{5.90} 
        & 65.39 \poschangenew{13.40} \\
        
        \hspace{2mm}OpenHands 
        & 10.38 \poschangenew{2.60} 
        & \;9.72 \poschangenew{3.89} 
        & 15.27 \poschangenew{0.37} 
        & 15.85 \negchangenew{0.48} 
        & 13.00 \poschangenew{5.85} 
        & 64.22 \poschangenew{12.23} \\
        
        \hspace{2mm}\ourmethod 
        & \textbf{12.20} \poschangenew{4.42} 
        & \textbf{11.63} \poschangenew{5.80} 
        & 15.33 \poschangenew{0.43} 
        & 16.15 \negchangenew{0.18} 
        & \textbf{14.75} \poschangenew{7.60} 
        & \textbf{70.06} \poschangenew{18.07} \\
        
        \bottomrule
        \end{tabular}
    }
    \label{tab:rq1-results}
\end{table*}

\section{Experimental Setup}
\subsection{Research Questions (RQs)}
We evaluate the effectiveness of \ourmethod across multiple dimensions, aiming to answer the following research questions:

\begin{itemize}[leftmargin=0.5cm]
\item \textbf{RQ1}: How effectively does \ourmethod answer code questions at repository scale? 

\item \textbf{RQ2}: To what extent do individual components influence the overall performance of our method?

\item \textbf{RQ3}: How does the number of exploration iterations in \ourmethod affect the answering performance? 

\item \textbf{RQ4}: What are the efficiency implications of \ourmethod while improving accuracy?

\item \textbf{RQ5}: How does \ourmethod perform across different types of repository-level code questions?

\end{itemize}
\subsection{Baselines and Models}
\label{sec:BaselinesandModels}

We evaluate \ourmethod against four categories of baselines: Direct Prompting, RAG-based Methods, Agent-based Methods, and Commercial Tools.

\textbf{Direct Prompting.} This baseline queries LLMs directly without providing any repository context, aiming to assess the intrinsic knowledge of the models.

\textbf{RAG-based Methods.} We evaluate two retrieval-augmented generation strategies:
\begin{itemize}[leftmargin=0.5cm]
    \item \textbf{Function Chunking RAG}~\cite{wang2024rlcoder}, which parses the repository into function- or class-level chunks as retrieval units. Each function or class is treated as an individual code entity, and a semantic index is constructed over these entities using code embeddings. At query time, the retriever uses \texttt{voyage-code-3} to encode both the query and code entities, and retrieves the top-10 most relevant chunks as context for answer generation.
    \item \textbf{Sliding Window RAG}~\cite{zhang2023repocoder}, which segments code files into overlapping chunks using a sliding window mechanism. Specifically, each file is divided into chunks of 500 lines with an overlap of 50 lines between adjacent windows. A semantic index is constructed over these chunks using \texttt{voyage-code-3} embeddings. At query time, the top-10 most relevant chunks are retrieved to provide context for answer generation.
\end{itemize}

\textbf{Agent-based Methods.} We adopt two representative ReAct-style code agents:
\begin{itemize}[leftmargin=0.5cm]

    \item \textbf{\textsc{SWE-Agent}}~\cite{yang2024swe}. We use SWE-agent v1.0, retaining all original tools and core agent logic, with only minor modifications to the entry and exit interfaces to support QA tasks.
    
    \item \textbf{\textsc{OpenHands}}~\cite{wang2024openhands}. We use version 1.1.0 of both \texttt{openhands-tools} and \texttt{openhands-sdk}, with the default agent ``\texttt{get\_default\_agent}'', which includes tools such as terminal, file editor, task tracker, finish, and think. The agent is adapted to the QA setting without modifying its core reasoning or tool-use mechanism.

\end{itemize}

For all agent-based methods, we set the maximum number of interaction iterations to 15. When the iteration limit is reached, the final response is generated based on the accumulated message history. For \ourmethod, only the adopted reasoning trajectory is used rather than all node messages.

\textbf{Commercial Tools.} We further compare against two state-of-the-art proprietary systems:
\begin{itemize}[leftmargin=0.5cm]
    \item \textbf{Tongyi Lingma} (VSCode plugin v2.5.16, auto mode)
    \item \textbf{Cursor-agent} (v2026.01.09-231024f, auto mode)
\end{itemize}

Tongyi Lingma relies on its proprietary model, while Cursor operates in its default ``auto'' mode, which dynamically selects the most suitable model based on the user query and integrates built-in retrieval and orchestration mechanisms. For both systems, we do not impose any restrictions on inference cost or the number of interaction turns. Instead, we directly adopt the final answer produced by the agent after its complete internal reasoning and tool usage process.

All open-source methods are evaluated with the same set of underlying LLMs to ensure a fair comparison, including four powerful and representative models: \textbf{GLM-4.6}~\cite{glm-4.6}, \textbf{Kimi K2}~\cite{kimi-k2-0905-preview}, \textbf{Qwen3-Coder-480B-A35B-Instruct}~\cite{qwen3-coder-480b-a35b-instruct}, and \textbf{GPT-5.1}~\cite{gpt-5.1-2025-11-13}.

\begin{table*}[t]
    \centering
    \small
    \renewcommand{\arraystretch}{1.05}
    \caption{Ablations on Key Components 
    }
    \resizebox{0.9\linewidth}{!}{
    \begin{tabular}{l@{\hspace{9mm}}c@{\hspace{3mm}}c@{\hspace{3mm}}c@{\hspace{3mm}}c@{\hspace{3mm}}c@{\hspace{3mm}}c}
        \toprule
        \multirow{2}[2]{*}{\textbf{Variant}} & \multicolumn{5}{c}{\textbf{Evaluation Metrics}} & \multirow{2}[2]{*}{\hspace{2mm} \textbf{Overall}} \\
        \cmidrule{2-6} 
        & \textbf{Correctness} & \textbf{Completeness} & \textbf{Relevance} & \textbf{Clarity} & \textbf{Reasoning} & \\ 
        \midrule
        \ourmethod & 10.97 & 10.17 & 13.82 & 16.19 & 13.28 & 64.43 \\
        
        \hspace{4mm} w/o MCTS 
        & 9.18 \negchangenew{1.79} 
        & 9.62 \negchangenew{0.55} 
        & 13.27 \negchangenew{0.55} 
        & 16.88 \poschangenew{0.69} 
        & 13.21 \negchangenew{0.07} 
        & 62.16 \negchangenew{2.27} \\
        
        \hspace{4mm} w/o Perception Agent 
        & 9.38 \negchangenew{1.59} 
        & 9.77 \negchangenew{0.40} 
        & 13.56 \negchangenew{0.26} 
        & 16.77 \poschangenew{0.58} 
        & 13.22 \negchangenew{0.06} 
        & 62.70 \negchangenew{1.73} \\
        
        \hspace{4mm} w/o Evaluation Agent 
        & 8.55 \negchangenew{2.42} 
        & 9.41 \negchangenew{0.76} 
        & 13.11 \negchangenew{0.71} 
        & 16.45 \poschangenew{0.26} 
        & 13.41 \poschangenew{0.13} 
        & 60.93 \negchangenew{3.50} \\
        
        \hspace{4mm} w/o Semantic Search 
        & 10.48 \negchangenew{0.49} 
        & 10.08 \negchangenew{0.09} 
        & 13.49 \negchangenew{0.33} 
        & 16.00 \negchangenew{0.19} 
        & 13.31 \poschangenew{0.03} 
        & 63.36 \negchangenew{1.07} \\
        \bottomrule
    \end{tabular}%
    }
    \label{tab:rq2-results}
\end{table*}

\subsection{Dataset}
\label{sec:dataset}
We evaluate all methods on the \ourdata benchmark~\cite{peng2026swe}. \ourdata is a repository-level code question answering benchmark designed to evaluate automated QA systems in realistic software development environments. The benchmark originally consists of 576 high-quality question-answer pairs drawn from a set of open-source Python repositories. To further improve its coverage and mitigate potential data leakage issues, three additional repositories---\texttt{conan}, \texttt{streamlink}, and \texttt{reflex}---were incorporated, selected from SWE-Bench Live. After this expansion, the benchmark comprises 15 repositories in total, containing 720 question-answer pairs.

\subsection{Evaluation Metrics}
\label{sec:EvaluationMetrics}
Following the evaluation protocol of \ourdata~\cite{peng2026swe}, we adopt the LLM-as-a-Judge paradigm to assess answer quality. To mitigate potential evaluation biases introduced by individual LLM judges, we further enhance the protocol by employing three distinct judge models, namely GPT-5.4~\cite{openai2025gpt54}, Claude-Sonnet-4-6~\cite{anthropic2026sonnet46}, and Gemini-3.1-Pro~\cite{gemini2026pro}. The final score is obtained by averaging the outputs of these judges. This multi-judge setup improves robustness and reduces the variance associated with any single evaluator. The LLM-as-a-judge prompt is provided in Appendix~\ref{sec:prompts}.

The evaluation is conducted across five key dimensions:
\begin{itemize}
    \item \textbf{Correctness}: assesses the factual accuracy of the answer.
    \item \textbf{Completeness}: evaluates how thoroughly the answer addresses all aspects of the question.
    \item \textbf{Relevance}: measures how well the answer matches the query.
    \item \textbf{Clarity}: reflects readability and ease of understanding.
    \item \textbf{Reasoning}: evaluates the logical soundness, depth, and validity of the reasoning process underlying the answer.
\end{itemize}

Each dimension is scored independently on a continuous scale from 0 to 20, following a detailed scoring rubric with predefined criteria for different score ranges. Specifically, each score interval (e.g., 16--20, 12--16, etc.) is associated with explicit qualitative guidelines that describe the expected level of performance for that dimension, enabling fine-grained and consistent evaluation across responses.


\subsection{Implementation Details}
Our \texttt{SemanticSearch} component employs the \texttt{voyage-code-3} code embedding model. For MCTS, we cap the number of children expanded per node at three (\textit{max\_expand} = 3) and set the maximum exploration iterations to 15 for the main results reported in Table~\ref{tab:rq1-results}. During the exploration phase of MCTS, we use a temperature of 0.7 to balance exploration and exploitation, while for answer generation we set the temperature to 0. For answer generation, we fix the temperature to 0 while keeping other decoding parameters (e.g., top\_p) at their default values as provided by each API, without additional tuning.

\section{Results}
\subsection{RQ1: Main Results}

Table~\ref{tab:rq1-results} compares the overall results by various methods.
Across models, \ourmethod is the strongest open-source system and closes the gap to commercial tools. With GPT-5.1 it achieves an overall score of 70.06, surpassing Tongyi Lingma at 69.12 and remaining close to Cursor (70.71). With GLM-4.6, Kimi K2 and Qwen3-Coder-480B-A35B-Instruct it remains the top open-source method, and on Kimi K2, further narrows the gap to commercial tools.

The most significant gains manifest in Correctness, Completeness, and Reasoning---dimensions critically important for multi-hop retrieval and evidence synthesis, while maintaining strong Relevance and fluent Clarity. This pattern reveals that MCTS-guided exploration and index-grounded actions excel at generating answers that are not only more accurate and comprehensive, but also more rigorously reasoned.

RAG variants outperform direct prompting but still trail agentic approaches, particularly on Correctness and Completeness, which indicates that RAG's retrieval may introduce substantial irrelevant code while missing critical code segments. \ourmethod strengthens the substantive dimensions across backbones and, with stronger backbones, reaches or exceeds commercial performance. Overall, systematic, feedback-guided search over repositories proves more effective than flat retrieval or single-trajectory agents. However, while SWE-agent and OpenHands execute more sophisticated actions, they may commit to incorrect search trajectories without timely correction mechanisms, causing them to persist in unproductive explorations, whereas our MCTS-guided approach enables continuous feedback-based course correction.    

\begin{tcolorbox}[
  colback=gray!5,
  colframe=black,
  boxrule=0.5pt, 
  title=Finding 1,
  fonttitle=\bfseries,
  left=2mm,
  right=2mm,
  separator sign dash
]\ourmethod achieves state-of-the-art performance among open-source systems and approaches leading commercial tools, especially in correctness, completeness, and reasoning.
\end{tcolorbox}

\subsection{RQ2: Ablation Study}
To understand the contribution of individual components, we conduct ablation studies removing: (i) MCTS framework (replaced with greedy single-trajectory search), (ii) Perception Agent (without state summarization), (iii) Evaluation Agent (reverted to rollout-based evaluation), and (iv) Semantic Search (exact name-based retrieval only). The results with Qwen3-Coder-480B-A35B-Instruct as the base model in Table~\ref{tab:rq2-results} demonstrate that each component meaningfully contributes to overall performance. Removing MCTS causes a 2.27-point performance drop, effectively reducing our method to OpenHands-level performance and underscoring the critical value of guided exploration over greedy single-trajectory search. Notably, the Evaluation Agent emerges as the most critical component---its removal results in the largest performance degradation of 3.50 points, reflecting its essential role in guiding path selection within the MCTS framework. The Perception Agent and Semantic Search provide additional improvements of 1.73 and 1.07 points respectively, demonstrating that comprehensive state summarization and semantic-aware retrieval collectively enhance answer quality across all dimensions. 

The MCTS framework's substantial impact across all evaluation dimensions confirms that systematic exploration-exploitation balance is fundamental for repository-level reasoning.

The component hierarchy reveals that systematic exploration mechanisms outweigh individual optimization techniques in importance. The notable impact of removing the Evaluation Agent validates our design choice to replace expensive rollouts with learned value estimation, while semantic search contributes balanced improvements across diverse question types. These findings demonstrate that principled algorithmic innovations (MCTS, value estimation, semantic actions) drive primary performance gains, providing clear guidance for future development priorities in repository-level QA systems.

\begin{tcolorbox}[
  colback=gray!5,
  colframe=black,
  boxrule=0.5pt, 
  title=Finding 2,
  fonttitle=\bfseries,
  left=2mm,
  right=2mm,  
  separator sign dash
]
All components in \ourmethod prove essential with no redundancy: each removal incurs performance loss ($-$1.07 to $-$3.50), validating the necessity of our integrated design.
\end{tcolorbox}

\begin{table*}[t]
    \centering
    \small
    \renewcommand{\arraystretch}{1.05}
    \caption{Effect of exploration budget (higher is better).
    }
    \begin{tabular*}{0.88\textwidth}{@{\extracolsep{\fill}}c c c c c c c}
        \toprule
        \multirow{2}[2]{*}{\textbf{Max Nodes}} & \multicolumn{5}{c}{\textbf{Evaluation Metrics}} & \multirow{2}[2]{*}{\hspace{2mm} \textbf{Overall}} \\
        \cmidrule{2-6} 
        & \textbf{Correctness} & \textbf{Completeness} & \textbf{Relevance} & \textbf{Clarity} & \textbf{Reasoning} & \\ 
        \midrule
        5   & 8.27 & 7.97 & 13.71 & 16.30 & 8.81 & 55.06 \\
        10  & 10.43 & 9.99 & 13.64 & 16.21 & 12.70 & 62.97 \\
        15  & 10.97 & 10.17 & 13.82 & 16.19 & 13.28 & 64.43 \\
        20  & 11.57 & 10.76 & 13.83 & 16.29 & 12.88 & 65.33 \\
        30  & 11.43 & 10.89 & 13.77 & 16.16 & 12.90 & 65.15 \\
        \bottomrule
    \end{tabular*}
    \label{tab:rq3-results}
\end{table*}

\subsection{RQ3: Impact of Exploration Iterations}

To address RQ3, we analyze the impact of varying the maximum number of exploration iterations on the performance of \ourmethod. The results with Qwen3-Coder-480B-A35B-Instruct as the base model are summarized in Table~\ref{tab:rq3-results}. We can observe that increasing the maximum number of explored nodes steadily improves performance: the overall score rises from 55.06 (5 nodes) to 62.97 (10 nodes, +7.91) and to 65.33 (20 nodes, +10.27). Correctness increases from 8.27 to 11.57 (+3.30), Completeness from 7.97 to 10.76 (+2.79), and Reasoning from 8.81 to 12.88 (+0.07). Notably, the largest performance gain occurs between 5 and 10 nodes, indicating that the MCTS framework requires a minimum number of iterations to fully realize its advantage. Within the experimental range, more iterations generally lead to higher scores, reflecting more thorough exploration; however, the incremental gains diminish at higher iteration counts, and performance between 20 and 30 nodes is observed to be almost identical. This suggests that users should balance the trade-off between computational cost and performance improvement when choosing the number of exploration iterations.


\begin{figure}[t]
    \centering 
    \includegraphics[width=0.95\linewidth]{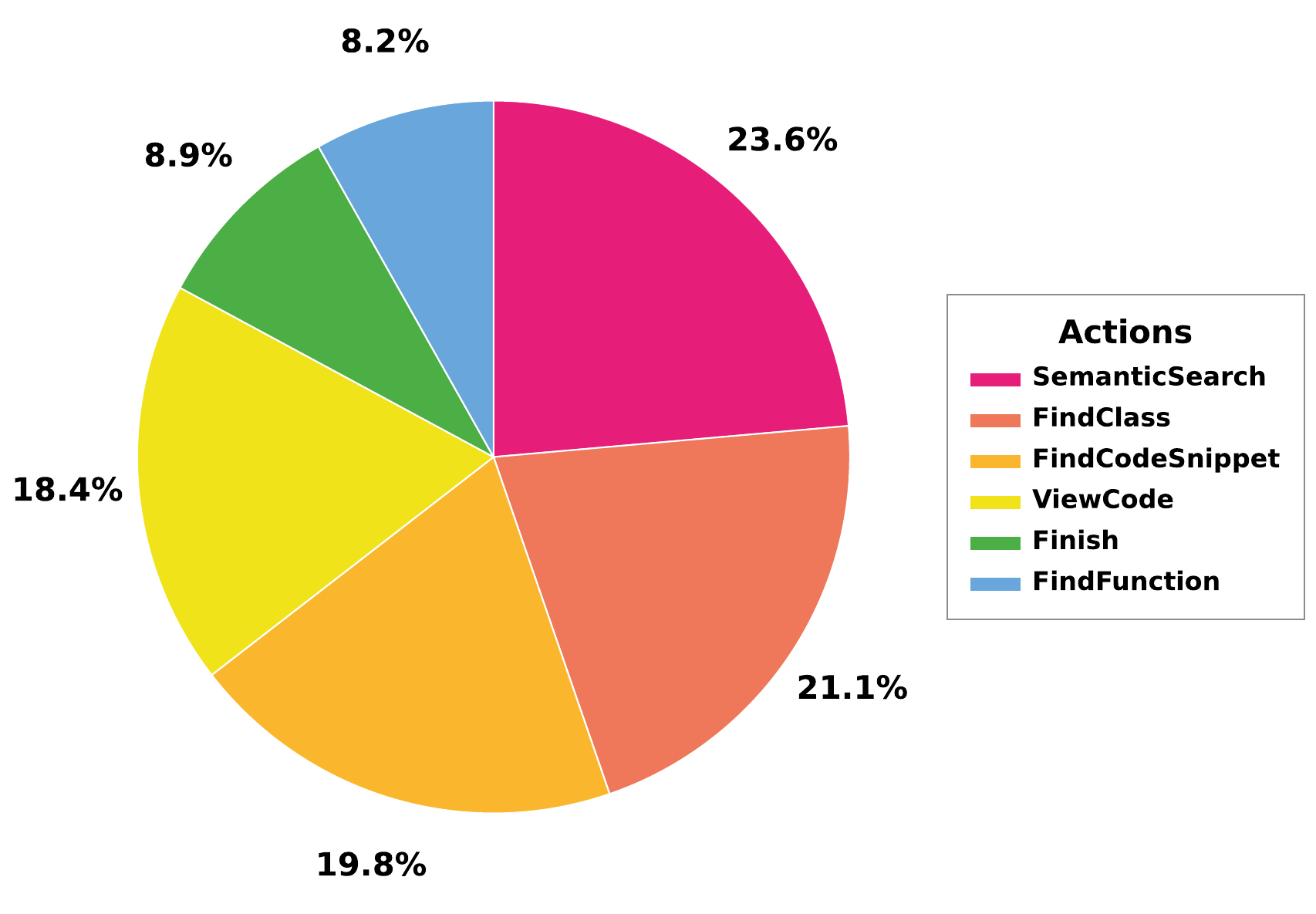}
    \caption{Distribution of action usage during trajectories.}
    \label{fig:action-usage-distribution}
\end{figure}

We further analyze the agent's exploration behavior by examining its action usage. The results are summarized in Figure~\ref{fig:action-usage-distribution}.

The distribution of actions used during trajectories (Figure~\ref{fig:action-usage-distribution}) clarifies the underlying mechanism. The agent's behavior is dominated by a combination of broad, meaning-based exploration (\texttt{Semantic\allowbreak Search}) and targeted, structure-aware navigation (\texttt{FindClass},\\ \texttt{FindCodeSnippet}), supplemented by frequent use of \texttt{ViewCode} for verification. Actions like \texttt{FindFunction} and the terminal \texttt{Finish} action are less common, indicating that the agent spends most of its budget actively gathering and confirming evidence. This pattern of guided exploration followed by verification directly contributes to the observed improvements in Correctness and Completeness as the exploration budget increases, as it allows the agent to build a more comprehensive and well-grounded understanding of the repository before synthesizing an answer.

\begin{tcolorbox}[
  colback=gray!5,
  colframe=black,
  boxrule=0.5pt, 
  title=Finding 3,
  fonttitle=\bfseries,
  left=2mm,
  right=2mm,  
  separator sign dash
]
Increasing the number of exploration iterations improves performance, but the gains diminish at higher iteration counts, with results plateauing at large budgets.
\end{tcolorbox}

\subsection{RQ4: Efficiency Analysis}

Appendix~\ref{app:token-usage} (Table~\ref{tab:token_cost}) summarizes the computational efficiency of all methods, including the number of input and output tokens. Under the same number of iterations (15), \ourmethod uses fewer input tokens compared to SWE-agent, OpenHands, and even commercial tools without iteration limits, while slightly more output tokens, mainly due to the design of the Perception Agent and Evaluation Agent. Overall, considering both input and output tokens, \ourmethod's computational cost is comparable to state-of-the-art agents and commercial tools, while still achieving better benchmark performance. These results highlight that \ourmethod achieves efficient token usage while delivering superior answer quality across diverse repositories.

\begin{tcolorbox}[
  colback=gray!5,
  colframe=black,
  boxrule=0.5pt, 
  title=Finding 4,
  fonttitle=\bfseries,
  left=2mm,
  right=2mm,  
  separator sign dash
]
\ourmethod achieves strong computational efficiency while delivering superior repository-level QA performance.
\end{tcolorbox}

\begin{table*}[t]
\centering
\small
\renewcommand{\arraystretch}{1.02}
\caption{Results by question type (higher is better).
}
\label{tab:type_results}
\begin{tabular*}{0.92\textwidth}{@{\extracolsep{\fill}}l c c c c c}
\toprule
\textbf{Question Type} &
\makecell{\textbf{GLM-4.6}} &
\makecell{\textbf{Kimi K2}} &
\makecell{\textbf{Qwen3-Coder} \\ \textbf{480B-A35B-Instruct}}  &
\makecell{\textbf{GPT-5.1}} &
\textbf{Average} \\ 
\midrule\rowcolor{gray!20}
What  & 63.67 & 67.12 & 63.30 & 67.89 & 65.50 \\ \hline
~Architecture exploration  & 59.80 & 62.64 & 60.49 & 63.41 & 61.59 \\
~Concept / Definition  & 66.80 & 71.00 & 65.66 & 70.62 & 68.52 \\
~Dependency tracing  & 64.41 & 67.72 & 63.75 & 69.64 & 66.38 \\
\hline\rowcolor{gray!20}
Why  & 68.01 & 68.79 & 65.10 & 71.49 & 68.35 \\ \hline
~Design rationale  & 70.02 & 70.26 & 66.28 & 73.95 & 70.13 \\
~Purpose Exploration & 68.84 & 69.46 & 65.83 & 73.11 & 69.31 \\
~Performance  & 65.17 & 66.65 & 63.19 & 67.41 & 65.61 \\
\hline\rowcolor{gray!20}
Where  & 65.03 & 66.21 & 65.05 & 69.58 & 66.47 \\ \hline
~Data / Control-flow  & 63.25 & 64.57 & 63.82 & 66.48 & 64.53 \\
~Feature Location  & 64.85 & 66.09 & 64.79 & 67.92 & 65.91 \\
~Identifier Location  & 66.99 & 67.97 & 66.54 & 74.34 & 68.96 \\
\hline\rowcolor{gray!20}
How  & 66.89 & 72.72 & 64.27 & 71.28 & 68.79 \\ \hline
~System Design  & 66.72 & 72.19 & 63.80 & 70.95 & 68.42 \\
~Algorithm Implementation  & 67.11 & 73.21 & 64.64 & 72.41 & 69.34 \\
~API / Framework Support  & 66.84 & 72.76 & 64.37 & 70.48 & 68.61 \\
\bottomrule
\end{tabular*}
\end{table*}

\subsection{RQ5: Performance across Question Types}

Table~\ref{tab:type_results} shows the performance of \ourmethod across different question types. The results indicate that \ourmethod performs strongly across all major question categories, with ``Why'' and ``How'' questions achieving the highest average scores, suggesting the model is particularly effective at understanding design rationale and reasoning about system design or algorithmic implementation. ``What'' and ``Where'' questions achieve slightly lower scores but still demonstrate solid performance, typically involving fetching definitions, basic concepts, or precise locations of features and identifiers. This detailed breakdown highlights the versatility of our approach while pointing to potential areas for improvement in handling these types of queries.

\begin{tcolorbox}[
  colback=gray!5,
  colframe=black,
  boxrule=0.5pt, 
  title=Finding 5,
  fonttitle=\bfseries,
  left=2mm,
  right=2mm,  
  separator sign dash
]
MCTS-guided, feedback-driven exploration substantially improves correctness,
completeness, and reasoning compared to both RAG and agentic baselines.
\end{tcolorbox}

\section{Discussion}

\subsection{Judge Reliability}
\label{sec:JudgeReliability}
To assess the reliability of our evaluation, we examine it from two
complementary angles: agreement \emph{among the three LLM judges}
(Section~\ref{sec:EvaluationMetrics}), and validation of the LLM panel
\emph{against human experts}.

Among the three judges, we measure ranking consistency with a
Spearman-style score whose definition and full derivation are given in
Appendix~\ref{app:judge-reliability}. We obtain an overall consistency of
$\rho_{\text{overall}} = 1$, indicating perfect agreement among the LLM
judges on system rankings.

We further validate the LLM panel against a panel of three human experts.
The LLM-consensus and human-consensus scores correlate strongly
(Pearson $r = 0.972$), and the two panels agree on $88.2\%$ of pairwise
system comparisons (Cohen's $\kappa = 0.725$, substantial). The complete
study---per-dimension Krippendorff's $\alpha$, per-answer human--LLM
correlations, panel-consensus agreement, and a length-bias
analysis---is reported in Appendix~\ref{app:judge-reliability}.

\subsection{Case Study}

To illustrate the differences in answer quality, we examine a
scikit-learn question: ``\textit{What does sklearn.\allowbreak
preprocessing.\allowbreak StandardScaler do to target variables?}''
(Figure~\ref{fig:case-study}). The baseline \textbf{OpenHands} retrieves
only snippets about how \texttt{StandardScaler} processes \emph{feature}
vectors, and therefore wrongly treats the target vector the same way---
missing that \texttt{StandardScaler} ignores the \texttt{y} argument of
\texttt{fit} entirely. In contrast, \textbf{\ourmethod} grounds its
answer in the actual API: it locates the \texttt{fit} documentation
(``\texttt{y: None, Ignored}''), notes that \texttt{transform} accepts
only \texttt{X}, and surfaces \texttt{TransformedTargetRegressor} as
sklearn's dedicated mechanism for target-variable transformation.

Besides, for a systematic account
of \emph{where} \ourmethod fails, Appendix~\ref{app:failure-modes} reports an error
analysis over the lowest-scoring trajectories, identifying four recurring
failure modes---wrong location (F1), adjacent-concept retrieval (F2),
keyword-only exploration (F3), and open-ended design (F4)---together with the
symbol-anchoring and early-retrieval patterns that characterize high-scoring
answers.

\subsection{Threats to Validity}
\subsubsection{Internal Validity}
A primary threat to internal validity is data contamination: LLMs may have
seen benchmark repositories or similar code during pre-training, achieving
high performance through memorization rather than genuine reasoning. We
mitigate this in three ways. First, we evaluate both open-weight and
proprietary models, reducing the chance that all models memorized the same
repositories (Section~\ref{sec:BaselinesandModels}). Second, we include a
retrieval-free baseline that reflects each model's unaided knowledge,
isolating gains from our exploration and reasoning mechanisms rather than
latent knowledge (Section~\ref{sec:BaselinesandModels}). Third, our dataset
draws on diverse, relatively less popular repositories, lowering the risk of
overlap with pre-training corpora (Section~\ref{sec:dataset}).

\subsubsection{External Validity}
First, in terms of model generalization, we evaluate four representative large models, covering both open-source and closed-source models, indicating that our findings are not limited to a single model type (Section~\ref{sec:BaselinesandModels}). Second to mitigate potential bias in evaluation, we employ three different models as judges, compute the average score, and verify the correlation among them, ensuring that the scoring is robust and reliable (Section~\ref{sec:JudgeReliability}). Third, to test \emph{language} generalization beyond Python, we evaluate \ourmethod on a Java subset (30 QA pairs over Strata, Fineract, and
Shiro). It ranks first among open-source methods and trails only the
commercial Cursor system, mirroring the Python ranking; full results are
reported in Appendix~\ref{app:java-results}.

\FloatBarrier
\section{Conclusion}
In this paper, we introduced \ourmethod, a novel agent-based framework that reformulates repository-level code question answering as a planning problem solved with Monte-Carlo Tree Search. Comprehensive evaluations demonstrate that \ourmethod achieves consistent 4-7\% improvements over SOTA agent baselines across multiple LLMs, with gains concentrated on Correctness, Completeness and Reasoning dimensions critical for multi-hop repository understanding. Ablation studies identify MCTS as the primary performance driver, while the design of Perception, Evaluation, and other agents also contributes to stability and efficiency. Our work shows the effectiveness of systematic tree
search for repository-level reasoning with solid efficiency and provides a foundation for building more capable software engineering assistants.

\section*{Data-Availability Statement}
Our code and data are available at \url{https://doi.org/10.5281/zenodo.21063159}.

\balance
\bibliography{ref}

\onecolumn
\appendix
\makeatletter
\@addtoreset{figure}{section}
\@addtoreset{table}{section}
\@addtoreset{equation}{section}
\makeatother
\renewcommand{\thefigure}{\thesection\arabic{figure}}
\renewcommand{\thetable}{\thesection\arabic{table}}
\renewcommand{\theequation}{\thesection\arabic{equation}}
\section{MCTS Workflow Illustration}
\label{app:mcts-workflow}

\begin{figure}[ht]
    \centering
    \includegraphics[width=0.85\textwidth]{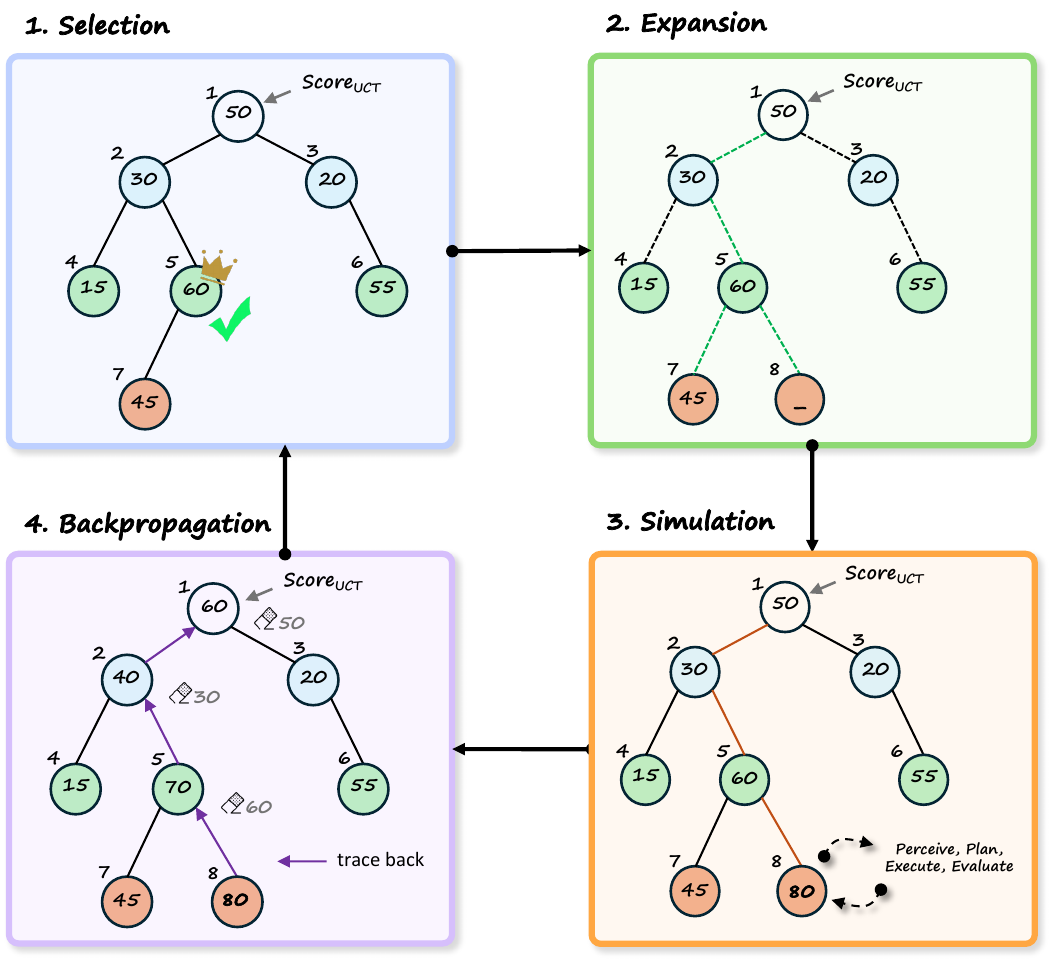}
    \caption{Illustration of our MCTS workflow.}
    \label{fig:mcts-workflow}
\end{figure}

\begin{figure}[p]
    \centering
    \includegraphics[width=0.70\textwidth,trim={0 70 510 70},clip]{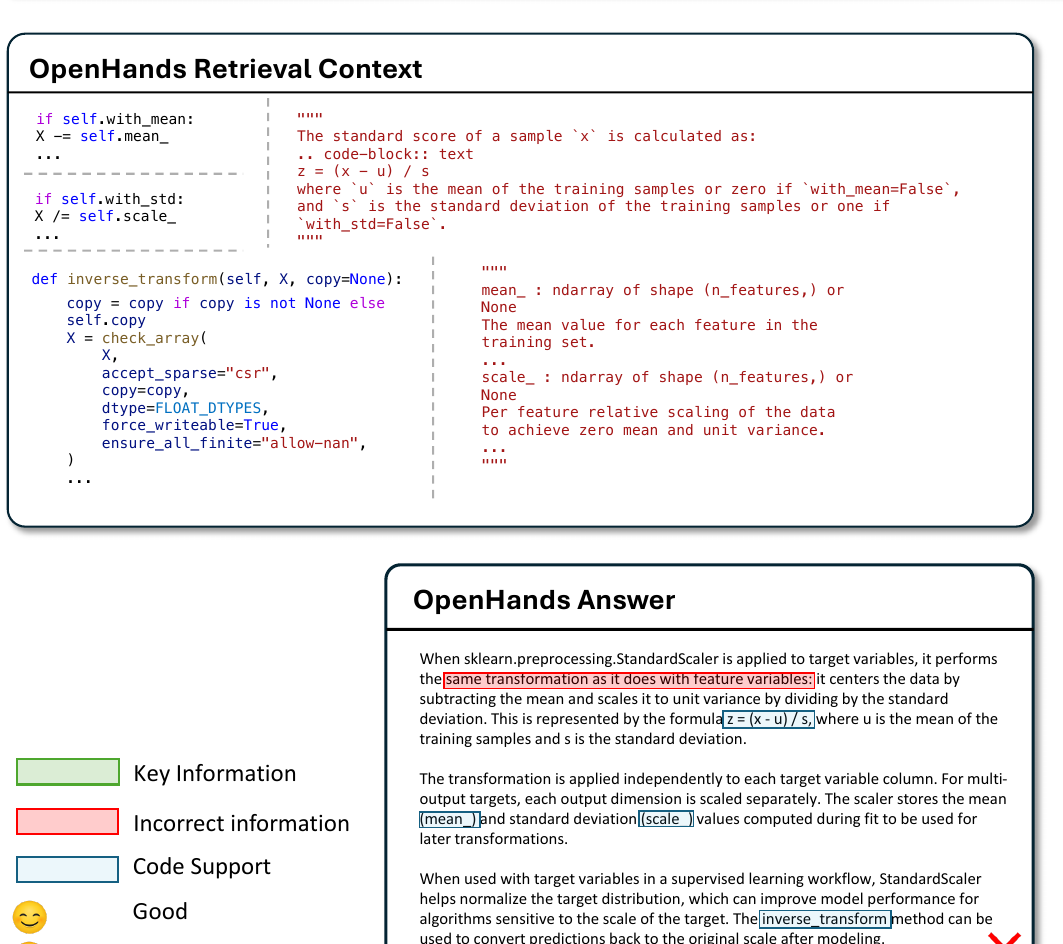}\\[-1mm]
    \textbf{(a) OpenHands: retrieved context and answer.}\\[2mm]
    \includegraphics[width=0.70\textwidth,trim={510 70 0 70},clip]{figs/casestudy.pdf}\\[-1mm]
    \textbf{(b) \ourmethod: retrieved context and answer.}
    \caption{Case study on scikit-learn \texttt{StandardScaler}, comparing the retrieved evidence and generated answers of OpenHands and \ourmethod.}
    \label{fig:case-study}
\end{figure}

\clearpage
\section{Token Usage}
\label{app:token-usage}

\begin{table}[ht]
\centering
\small
\setlength{\tabcolsep}{5pt}
\renewcommand{\arraystretch}{1.05}
\caption{Token usage per question, reported as input/output tokens (lower is better).}
\begin{tabular*}{0.96\textwidth}{@{\extracolsep{\fill}}l c c c c c}
\toprule
\textbf{Model} &
\makecell{\textbf{Qwen3-Coder-}\\\textbf{480B-A35B-Instruct}} &
\makecell{\textbf{Kimi K2}} &
\makecell{\textbf{GLM-4.6}} &
\makecell{\textbf{GPT-5.1}} &
\textbf{Average} \\
\cmidrule(lr){2-6}

Direct
& 94/65 & 94/34 & 94/76 & 94/87 & 94/63 \\

Function Chunking RAG
& 5,084/106 & 5,084/92 & 5,084/240 & 5,084/196 & 5,084/130 \\

Sliding Window RAG
& 6,302/112 & 6,302/110 & 6,302/233 & 6,302/201 & 6,302/138 \\

SWE-agent
& 165,206/4,546 & 154,295/1,398 & 70,870/1,902 & 122,833/6,100 & 126,026/4,627 \\

OpenHands
& 108,703/2,128 & 101,594/1,787 & 42,433/2,158 & 69,192/2,127 & 87,045/1,930 \\

\ourmethod
& 119,633/3,812 & 68,679/4,059 & 42,888/9,867 & 83,526/7,251 & 78,681/6,247 \\

\midrule

Tongyi Lingma
& \multicolumn{4}{c}{116,372/1,589}
& 116,372/1,589 \\

Cursor-agent
& \multicolumn{4}{c}{129,458/1,635}
& 129,458/1,635 \\

\bottomrule
\end{tabular*}
\label{tab:token_cost}
\end{table}

\def\DeepRepoQASupplementInput{}
\ifdefined\DeepRepoQASupplementInput
\else
\documentclass{article}

\usepackage[utf8]{inputenc}
\usepackage[T1]{fontenc}
\usepackage{lmodern}
\usepackage{booktabs}
\usepackage{amsmath}
\usepackage{geometry}
\geometry{a4paper, margin=1in}

\usepackage{tcolorbox}
\tcbuselibrary{breakable, skins, listings}

\usepackage{fancyvrb}   
\usepackage{xcolor}

\usepackage{hyperref}
\usepackage{tabularx}
\newcommand{\ourmethod}{\textsc{DeepRepoQA}}

\newcounter{prompt}

\begin{document}

\title{DeepRepoQA Supplementary Material}
\maketitle
\appendix
\fi
\section{LLM-as-a-Judge Reliability}
\label{app:judge-reliability}

All human and LLM evaluations in this section are conducted on answers generated by the \textbf{GPT-5.1 backbone} (60 questions \ensuremath{\times} 6 systems = 360 answers).

The 60 questions are sampled from the 15 evaluation repositories with stratification over the four top-level intent categories (\texttt{why}, \texttt{where}, \texttt{how}, \texttt{what}): for each repository we randomly draw one question from each category, yielding $15 \times 4 = 60$ questions. This design guarantees that every repository and every intent category is equally represented, so the reliability statistics below are not dominated by any single repository or question type.

\subsection{Intra-panel consistency --- Krippendorff's \ensuremath{\alpha} (interval level)}

\begin{table}[h]
\centering
\small
\caption{Intra-panel inter-rater reliability (Krippendorff's \ensuremath{\alpha}, interval level) for the human panel, the LLM panel, and the combined 6-rater panel, broken down by evaluation dimension.}
\begin{tabular}{lccc}
\toprule
Dimension & Human \ensuremath{\alpha} & LLM \ensuremath{\alpha} & All-6 \ensuremath{\alpha} \\
\midrule
Correctness & \textbf{0.837} & \textbf{0.908} & 0.878 \\
Completeness & \textbf{0.822} & \textbf{0.883} & 0.855 \\
Relevance & \textbf{0.828} & \textbf{0.832} & 0.838 \\
Clarity & 0.324 & 0.406 & 0.420 \\
Reasoning & \textbf{0.821} & \textbf{0.825} & 0.834 \\
\textbf{Overall} & \textbf{0.840} & \textbf{0.871} & \textbf{0.861} \\
\bottomrule
\end{tabular}
\end{table}

4/5 dimensions exceed the 0.80 substantial-reliability threshold for both panels. Clarity is consistently the hardest dimension for both human and LLM panels, indicating a domain-inherent difficulty.

\subsection{Individual cross-group Spearman \ensuremath{\rho} matrix (per-answer, all p \textless{} 0.001)}

\begin{table}[h]
\centering
\small
\caption{Per-answer Spearman \ensuremath{\rho} between each LLM judge and each human expert (all p \textless{} 0.001).}
\begin{tabular}{lcccc}
\toprule
 & Expert 1 & Expert 2 & Expert 3 & LLM row mean \\
\midrule
GPT-5.4 & 0.871 & 0.793 & 0.791 & 0.818 \\
Claude-Sonnet-4-6 & 0.661 & 0.915 & 0.918 & 0.831 \\
Gemini-3.1-Pro & 0.789 & 0.848 & 0.847 & 0.828 \\
\textbf{Human col mean} & 0.774 & 0.852 & 0.852 & \textbf{Cross-group mean: 0.826} \\
\bottomrule
\end{tabular}
\end{table}

Within-human pairwise \ensuremath{\rho}: E1--E2 = 0.680, E1--E3 = 0.668, E2--E3 = 0.917 \ensuremath{\rightarrow} \textbf{mean 0.755}.

The lower Claude--Expert1 entry (0.661) mirrors Expert 1's own low within-human correlations (0.668--0.680): Expert 1 also correlates weakly with the other two human experts, pointing to an idiosyncratic rating style specific to this expert.

\subsection{Panel-consensus agreement (mean-of-3 vs mean-of-3)}

Beyond individual rater consistency, we compare the two panels at the consensus level by averaging the three human ratings into a human-consensus score and the three LLM ratings into an LLM-consensus score. Table~\ref{tab:s13} reports the agreement between these two consensus signals on the 0--100 scale.

\begin{table}[h]
\centering
\small
\caption{Panel-level agreement between the human-consensus score (mean of 3 experts) and the LLM-consensus score (mean of 3 LLM judges).}
\label{tab:s13}
\begin{tabular}{lc}
\toprule
Measure & Value \\
\midrule
Pearson r & \textbf{0.972} {[}95\% CI 0.964--0.979, bootstrap 2,000 resamples{]} \\
Spearman \ensuremath{\rho} & \textbf{0.942} \\
MAE (0--100 scale) & \textbf{5.22 pts} (5.2\% of scale; inter-method gap \textasciitilde25
pts) \\
\bottomrule
\end{tabular}
\end{table}

MAE (5.22 pts) is roughly one-fifth of the typical gap between competing systems (\textasciitilde25 pts), so the panel-level signal is precise enough to distinguish methods reliably. Combined with the very high Pearson r (0.972) and Spearman \ensuremath{\rho} (0.942), the LLM-consensus closely tracks the human-consensus both in absolute level and in ordering.

\subsection{Pairwise preference agreement}

The question of primary practical interest is ``which system is better?'' Table~\ref{tab:s14} therefore reports agreement at the \emph{decision} level, converting each pair of system scores into a preference (A \textgreater{} B, tie, or B \textgreater{} A) and measuring how often the two panels reach the same verdict.

\begin{table}[h]
\centering
\small
\caption{Pairwise preference agreement between human and LLM panels on system-vs-system comparisons.}
\label{tab:s14}
\begin{tabular}{lc}
\toprule
Measure & Value \\
\midrule
Preference agreement rate & \textbf{88.2\%} \\
Cohen's \ensuremath{\kappa} (3-class: A \textgreater{} B / tie / B \textgreater{} A) & \textbf{0.725} (substantial) \\
System-level ranking Spearman \ensuremath{\rho} & \textbf{0.900} (p = 0.037) \\
\bottomrule
\end{tabular}
\end{table}

\subsection{Length-bias analysis}

We also examine whether answer length influences LLM scores independently of quality, using two analyses: an OLS regression that partials out system and dimension effects, and a stratified comparison across short / medium / long answers (Table~\ref{tab:s15}).

OLS model: \texttt{score\ \textasciitilde{}\ length\ +\ system\ +\ dim}.

Correlation of (LLM consensus \ensuremath{-} Human consensus) with answer length: r = \textbf{0.179}, p = 0.006, r\textsuperscript{2} = \textbf{0.032}.

Stratified human-LLM agreement and score difference:

\begin{table}[h]
\centering
\small
\caption{Length-stratified human-LLM agreement, showing the share of answers in each bin, the mean LLM\,\ensuremath{-}\,Human score gap, and the per-answer Pearson r between the two panels.}
\label{tab:s15}
\begin{tabular}{lccc}
\toprule
Length bin & percentage & Mean (LLM \ensuremath{-} Human) & Pearson r (human vs LLM) \\
\midrule
Short (\textless{} 100 words) & 53.9\% & \textbf{\ensuremath{-}4.85} & \textbf{0.983} \\
Medium (100--300 words) & 14.6\% & \textbf{\ensuremath{-}1.16} & \textbf{0.985} \\
Long (\textgreater{} 300 words) & 31.4\% & \textbf{\ensuremath{-}3.06} & \textbf{0.919} \\
\bottomrule
\end{tabular}
\end{table}

LLM scores are uniformly \emph{lower} than human scores across all length bins (LLMs are more stringent). Human-LLM agreement is uniformly excellent (r \ensuremath{\geq} 0.92).

\subsection{Inter-judge ranking consistency}

To assess the reliability of our evaluation, we also measure the ranking consistency \emph{among the three LLM judges}. We observe a high level of consistency which, combined with the human verification protocol of SWE-QA, indicates that the judges are reliable. The computation is as follows.

Let $R_{i,j}(m)$ denote the rank assigned by judge $j$ to baseline $m$ for target model $i$, and let $S_{j,k}=\sum_{m=1}^{n}\left(R_{i,j}(m)-R_{i,k}(m)\right)^2$. The per-model consistency score is
\begin{equation}
\rho_i = \frac{1}{3} \sum_{(j,k)\in \{(1,2),(1,3),(2,3)\}}
\left(1 - \frac{6 S_{j,k}}{n(n^2-1)}\right),
\end{equation}
and the overall consistency is $\rho_{\text{overall}} = \frac{1}{M} \sum_{i=1}^{M} \rho_i$. In our experiments, we find $\rho_{\text{overall}} = 1$, indicating perfect agreement among the LLM judges; this judge-level agreement is complemented by the human-validation study reported in the preceding subsections.

\bigskip

\section{Cross-language Generalization --- Java Results}
\label{app:java-results}
\begin{table}[h]
\centering
\small
\caption{Cross-language generalization on the Java subset (30 QA pairs across Strata, Fineract, and Shiro). DeepRepoQA ranks first among open-source methods and trails only the commercial Cursor system, mirroring the Python ranking.}
\begin{tabular}{lcccccc}
\toprule
Method & Correctness & Completeness & Relevance & Clarity & Reasoning & Overall \\
\midrule
Direct & 5.42 & 4.18 & 10.85 & 11.36 & 5.71 & 37.52 \\
Function-Chunking RAG & 7.83 & 6.97 & 12.14 & 12.05 & 7.62 & 46.61 \\
Sliding-Window RAG & 8.46 & 7.51 & 12.48 & 12.23 & 8.05 & 48.73 \\
SWE-agent & 11.27 & 9.94 & 13.78 & 12.58 & 11.32 & 58.89 \\
OpenHands & 11.41 & 10.08 & 13.91 & 12.65 & 11.46 & 59.51 \\
\textbf{DeepRepoQA} & \textbf{12.29} & \textbf{11.30} & \textbf{14.37} & \textbf{13.13} & \textbf{12.21} & \textbf{63.29} \\
Tongyi Lingma (commercial) & 11.84 & 10.51 & 14.08 & 12.79 & 11.79 & 61.01 \\
Cursor (commercial) & 12.85 & 11.74 & 14.62 & 13.31 & 12.59 & 65.10 \\
\bottomrule
\end{tabular}
\end{table}

\textbf{Dataset}: SWE-QA Java subset --- Strata (quantitative finance), Fineract (banking infrastructure), Shiro (security framework); 30 QA pairs total, same evaluation protocol as Python main experiments.

DeepRepoQA ranks first among open-source methods (+3.78 over OpenHands), outperforms Tongyi Lingma (+2.28), and trails Cursor by 1.81 pts --- consistent with the Python setting.

\bigskip

\section{Failure Mode Case Studies}
\label{app:failure-modes}

\subsection{Handling of misleading retrievals}

When \texttt{SemanticSearch} returns a misleading snippet, the Evaluation Agent in most cases recognises that the retrieved evidence is unhelpful and assigns a low score, which triggers backtracking via UCT. If the Evaluation Agent incorrectly assigns a high score, the trajectory continues and is likely to accumulate further unhelpful evidence, eventually triggering a delayed backtrack once the cumulative value signal falls below that of an unexpanded sibling. In a minority of cases this two-stage correction mechanism fails to recover, and the trajectory terminates on the wrong path --- the failure modes catalogued below characterise when and why this occurs.

\subsection{Failure modes}

From a manual inspection of the lowest-scoring trajectories across all question categories, we identify four recurring failure modes (F1--F4). Table~\ref{tab:s52} lists each mode together with the question categories it most affects and its underlying root cause; Sections~\ref{sec:f1}-\ref{sec:f4} then walk through a representative case for each mode.

\begin{table}[h]
\centering
\small
\caption{Taxonomy of the four observed failure modes (F1--F4), the question categories they most affect, and their root causes.}
\label{tab:s52}
\begin{tabularx}{\linewidth}{l X X}
\toprule
Mode & Categories most affected & Root cause \\
\midrule
F1 Wrong location & where/iden-loca, where/funct-loca & Semantic \ensuremath{\neq} literal match; no verification step \\
F2 Adjacent concept & why/purpose, where/data-control-flow & Semantic ambiguity causes all branches to share wrong initial retrieval \\
F3 Keyword-only exploration & why/design-rationale, how/algo-impl & No structural navigation fallback when semantic search stalls \\
F4 Open-ended design & how/system-design & Multiple valid answers; reference-matching rubric penalises divergent but correct designs \\
\bottomrule
\end{tabularx}
\end{table}

\subsection{Success patterns}

Three factors consistently distinguish high-scoring answers from failures. First, \textbf{early retrieval precision}: 100\% of top-scoring cases (score \ensuremath{\geq} 73) reached the correct file within the first two hops; cases requiring 15+ node expansions almost never recover from an incorrect initial retrieval. Second, \textbf{symbol-level anchoring}: 80\% of top-scoring cases used \texttt{FindClass} or \texttt{FindFunction} with an explicit \texttt{file\_pattern} to pin the answer to a specific code span, preventing hallucination of adjacent code. Third, \textbf{\texttt{ViewCode} immediately before \texttt{Finish}}: re-reading the actual source one more time before drafting reduces paraphrase drift and is strongly correlated with high Correctness scores. Conversely, \textbf{Clarity is high even when Correctness is low} (Clarity mean 15.1 vs.~Correctness mean 10.5 in the full evaluation), confirming that fluent prose does not imply factual correctness --- Correctness and Completeness are the meaningful quality signals.

\subsection{\texorpdfstring{Representative success case --- \texttt{why/purpose}}{Representative success case --- why/purpose}}

\textbf{Question}: \emph{Why does the exception class raised when a computed variable shadows a base variable inherit from both the framework's base exception class and the built-in Python exception for naming conflicts?}

\textbf{Winning path} (8 nodes, 4 steps):
1. \texttt{SemanticSearch("exception\ class\ for\ computed\ variable\ shadowing\ a\ base\ var")}
2. \texttt{FindClass(ComputedVarShadowsBaseVarsError,\ file=reflex/utils/exceptions.py)}
3. \texttt{ViewCode(\textless{}that\ file\textgreater{})}
4. \texttt{Finish}

\textbf{Why it worked}: a natural-language description of a concrete entity mapped cleanly to a single class definition; \texttt{SemanticSearch} returned the right file on the first hop; \texttt{ViewCode} provided the docstring cited verbatim in the answer. The answer was generated from \textbf{read code + docstring}, not LLM inference, which is what kept Correctness high (17/20).

\subsection{\texorpdfstring{Representative failure case F1 --- \texttt{where/iden-loca}}{ Representative failure case F1 --- where/iden-loca}}
\label{sec:f1}
\textbf{Question}: \emph{Where is the function that parses argument specifications called to build the argument list for an event handler?}

\begin{table}[h]
\centering
\small
\caption{Failure case F1 (wrong location). The reference identifier \texttt{parse\_args\_spec} is replaced by a semantically similar but literally distinct function, with no verification step.}
\begin{tabularx}{\linewidth}{l X X}
\toprule
 & Reference & Candidate \\
\midrule
Location & \texttt{parse\_args\_spec} at \texttt{reflex/event.py:2085} & \texttt{\_extract\_func\_kwargs\_as\_ast\_nodes} in \texttt{pyi\_generator.py} \\
Scores & --- & Correctness 2, Completeness 1, Relevance 2 \\
\bottomrule
\end{tabularx}
\end{table}

\textbf{Root cause}: \texttt{SemanticSearch} returned a semantically similar but literally wrong function (\texttt{\_extract\_func\_kwargs\_as\_ast\_nodes} also parses kwargs but in a different file and context). The agent stopped after a single \texttt{ViewCode} and called \texttt{Finish} with high confidence, never attempting \texttt{FindFunction(parse\_args\_spec)} to verify the literal identifier. For \texttt{where}-type questions, the agent lacks a verification step that checks whether the question's literal identifier appears in the retrieved code.

\subsection{\texorpdfstring{Representative failure case F2 --- \texttt{why/purpose}}{Representative failure case F2 --- why/purpose}}

\textbf{Question}: \emph{Why is the application instance and its module encapsulated in a structured container within the framework?}

\begin{table}[h]
\centering
\small
\caption{Failure case F2 (adjacent concept). The ambiguous phrase ``structured container'' steers retrieval toward \texttt{AppWrap} (a UI wrapper) instead of the reference target \texttt{AppInfo} (a data container).}
\begin{tabularx}{\linewidth}{l X X}
\toprule
 & Reference & Candidate \\
\midrule
Target & \texttt{AppInfo} (a \texttt{NamedTuple} holding \texttt{App} + its module) & \texttt{AppWrap} (a component wrapping the root rendered tree) \\
Scores & --- & Correctness 4, Completeness 3, Relevance 5 \\
\bottomrule
\end{tabularx}
\end{table}

\textbf{Root cause}: the phrase ``structured container'' is semantically ambiguous --- it applies to both \texttt{AppInfo} (a data container) and \texttt{AppWrap} (a UI wrapper component). \texttt{SemanticSearch} surfaced \texttt{AppWrap} because that token appears more frequently in documentation. The agent committed to this interpretation across all 17 MCTS nodes without reconsidering, illustrating how semantic ambiguity in the question can cause all branches to share the same incorrect initial retrieval, making backtracking ineffective.

\subsection{\texorpdfstring{Representative failure case F3 --- \texttt{why/design-rationale}}{Representative failure case F3 --- why/design-rationale}}

\textbf{Question}: \emph{Why does the design constraint prioritize framework infrastructure hooks before other hook categories in the code generation process?}

\textbf{What went wrong}: over 7 steps the agent issued repeated \texttt{SemanticSearch} and \texttt{FindCodeSnippet} calls with progressively varied keywords (``hook category sort key'', ``order compiler code generation'', etc.), hitting the 20-node cap at 332 s. It never discovered the \texttt{HookPosition} enum or the actual ordering function. The agent had no fallback when keyword search stalled: it did not inspect directory structure, did not follow imports, and did not switch to structural navigation --- it only varied keywords.

\subsection{\texorpdfstring{Representative failure case F4 --- \texttt{how/system-design}}{Representative failure case F4 --- how/system-design}}
\label{sec:f4}
\textbf{Question}: \emph{How can the collapsible content panel component be redesigned to support runtime theming changes without breaking existing usage?}

\begin{table}[h]
\centering
\small
\caption{Failure case F4 (open-ended design). The agent produces a technically sound alternative redesign, but the reference-based LLM judge penalises divergence from the single prescribed solution.}
\begin{tabularx}{\linewidth}{l X X}
\toprule
 & Reference & Candidate \\
\midrule
Proposed solution & Specific approach involving theme token propagation via CSS variables & Alternative design using \texttt{add\_style()} and inherited data attributes \\
Scores & --- & Correctness 7, Completeness 8, Relevance 9 \\
\bottomrule
\end{tabularx}
\end{table}

\textbf{Root cause}: this is a \emph{redesign} question where the reference prescribes one specific engineering solution. The agent produced a plausible and technically sound alternative design, but the LLM judge penalised the divergence from the reference's specific proposal. This is partly an \textbf{evaluation artefact} --- LLM-as-judge compares against a single reference answer --- and partly an inherent limitation of open-ended design questions: code retrieval provides little guidance on which of many valid solutions is ``correct''. Unlike localization or explanation questions, no amount of additional retrieval steps would have led the agent to the reference's specific solution.

\bigskip

\section{Prompts}\label{sec:prompts}

\begin{tcolorbox}[colframe=black, colback=white,
    coltitle=white, colbacktitle=black,
    title=Prompt Template for Planning,
    boxrule=0.8pt,
    fonttitle=\mdseries\small,
    fontupper=\ttfamily\scriptsize,
    enhanced,
    rounded corners,
    listing only,
    breakable
]
\refstepcounter{prompt}
\label{prompt1}
\begin{Verbatim}
You are an autonomous AI assistant with superior skills in answering questions of a 
repository.
You need to answer the question exactly based on the information you can get from the 
available functions.
As you're working autonomously, you cannot communicate with the user but must rely on 
information you can get from the available functions.

# Action and ReAct Guidelines

1. Analysis First
   - Review all previous actions and their observations
   - Understand what has been done and what information you have

2. Document Your Thoughts
   - ALWAYS write your reasoning in <thoughts> tags before any action
   - Explain what you learned from previous observations
   - Justify why you're choosing the next action
   - Describe what you expect to learn/achieve

3. Single Action Execution
   - Run ONLY ONE action at a time
   - Choose from the available functions
   - Never try to execute multiple actions at once

4. Wait and Observe
   - After executing an action, STOP
   - Wait for the observation (result) to be returned
   - Do not plan or execute any further actions until you receive the observation

# Workflow Overview

1. Understand the Question
   * Review the Question: Carefully read the question provided in <task>
   * Identify Needed Information: Analyze the question to determine what parts of the 
   codebase you need to understand
   * Plan Your Investigation: Determine what components, functions, or classes you need
   to explore to find a complete answer

2. Locate Relevant Code
   * Primary Method - Search Functions:
       - FindClass - Search for class definitions by class name
       - FindFunction - Search for function definitions by function name
       - FindCodeSnippet - Search for specific code patterns or text
       - SemanticSearch - Search code by semantic meaning or natural language description
   * Secondary Method - ViewCode: Only use when you need to see:
       - Additional context not returned by searches
       - Specific line ranges from search results
       - Code referenced in other parts of the codebase

3. Gather Complete Information
   - Continue searching and viewing code until you have all necessary information
   - Investigate all relevant parts of the codebase
   - Look for related functionality that might be important

4. Analyze and Formulate Answer
   - Review all gathered information
   - Organize findings to create a complete and accurate answer
   - Ensure the answer is based on actual code, not assumptions

5. Complete Task
   - Use Finish you have sufficient information to provide a complete and accurate answer
   - Reference specific parts of the code in your final answer
   - Make sure the answer addresses all aspects of the question

# Important Guidelines

- Focus on the Specific Question
  - Answer exactly as asked, based on the code in the repository
  - Provide complete and accurate information
  - Do not assume code you haven't seen

- Code Context and Information
  - Base answers only on code you can see through searches and ViewCode
  - Use ViewCode to examine more code if needed
  - Reference specific code sections for clarity

- Task Completion
  - Finish only after gathering sufficient information
  - Cite specific evidence from the code
  - Ensure all relevant code has been explored

- State Management
  - Keep detailed records of viewed code and actions taken
  - Check history before performing a new action to avoid repetition
  - Use gathered information to inform next steps

# Additional Notes

- Think Step by Step
  - Document reasoning and thought process in <thoughts>
  - Build upon previous steps without unnecessary repetition

- Never Guess
  - Do not guess line numbers or code content
  - Use ViewCode to examine code when needed

# Examples

Here are some examples of how to use the available actions:

**Example 1: 
Task: I need to see the implementation of the DatabaseManager class to understand how 
it handles transactions
   <thoughts>To examine how the DatabaseManager class handles transactions, we need to 
   locate its implementation in the codebase.</thoughts>
   {"tool": "FindClass", "file_pattern": null, "class_name": "DatabaseManager"}

**Example 2: 
Task: Find the calculate_interest function in our financial module to review its logic
   <thoughts>To review the logic of the calculate_interest function, we need to locate 
   its implementation in the financial module.</thoughts>
   {"tool": "FindFunction", "file_pattern": "financial/**/*.py", "function_name": 
   "calculate_interest", "class_name": null}

*Note: This is a condensed version for display purposes. The actual prompt contains 
complete examples for all action tools.*

\end{Verbatim}
\end{tcolorbox}

\begin{tcolorbox}[colframe=black, colback=white, 
    coltitle=white, colbacktitle=black,
    title=Prompt Template for Perceiving and Generating Feedback,
    boxrule=0.8pt,
    fonttitle=\mdseries\small,
    fontupper=\ttfamily\scriptsize,
    enhanced,
    rounded corners,
    listing only,
    breakable,
]
\refstepcounter{prompt}
\label{prompt2}
\begin{verbatim}
You are a feedback agent that guides an AI assistant's next action.

---
##  CRITICAL: ACTION AGENT LIMITATIONS

The action agent receiving your feedback:

- CANNOT see the search tree  
- Has NO CONTEXT about node relationships  
- Only knows about actions in its direct trajectory  
- Cannot understand references to nodes without proper context  
- Is at a new node that has NO ACTION YET --- it needs your guidance for what to do next  

---
##  REQUIRED FEEDBACK STRUCTURE

### 1. CURRENT NODE CONTEXT
You must start by describing:

- **Position in tree:** `"You are at Node X, which is [position relative to root]"`
- **Current state:** `"Your node is currently empty and awaiting your first action"`
- **Parent context:** `"Your parent node (Node Y) [describe what parent did]"`
- **Relationship to solutions:** `"There are [N] terminal nodes in [relationship] 
branches"`

> **Note:** The current node is ALWAYS empty and awaiting its first action --- never 
describe it as having done something already.

--
##  CORRECT EXAMPLES

**Current Node Context Example:**
"You are at Node 8, which is your first action from the root.
Your node is currently empty and awaiting your first action.
Your parent (Node 1) performed a FindCodeSnippet action that didn't add new context.
There are three terminal nodes in parallel branches (Nodes 7, 9, and 14) that have
reached finished states with different approaches."


---
##  INCORRECT EXAMPLES --- DO NOT USE

- `"Node 8 is empty and expandable"`  
- `"The current node needs to explore improvements"`  
- `"We should validate the existing solution"`  
- Any description implying the current node has already taken an action  

---
##  INPUT STRUCTURE

1. **Tree Visualization:** ASCII representation showing:  
   - Node IDs and relationships  
   - Action types at each node  
   - Key outcomes and observations  

2. **Original Task:** The problem to solve  

3. **Message History:** Chain of executed actions leading to current state  

4. **Tree Structure:**  
   - **Parent Node:** Your current starting point, the last successfully executed 
   action  
   - **Current Node:** Your branch from the parent, waiting for your next action  
   - **Sibling Nodes:** Other independent solution attempts branching from the same 
   parent  
     (These are from different trajectories and have not happened in your current path)  

5. **Alternative Node Information:**  
   - Proposed actions and parameters  
   - Outcomes (from separate, independent trajectories)  
   - Warning flags for previously attempted approaches  

---
##  YOUR TASK

1. **Analyze the situation:**  
   - Start with current node context (position, state, parent, solutions)  
   - Consider sibling attempts (alternative universes)  
   - Learn from outcomes to avoid repeating unsuccessful approaches  
   - Contextualize feedback based on tree structure  
   - Always explain node relationships and attempts  
   - Inform about alternative approaches tried (files, tests, git diffs)  

2. **Suggest next action:**  
   - Clear, actionable guidance focusing on code query actions  
   - Based on lessons from other attempts  
   - Avoid repeating failed approaches  

3. **Optionally suggest node to expand:**  
   - Explain why this node is promising  
   - Leave as null if no strong preference  

> Remember: Focus on code query actions that help the agent better understand the 
codebase. Always provide proper context since the agent cannot see the tree.

**Important:** Please provide the answer in JSON format.

\end{verbatim}
\end{tcolorbox}

\begin{tcolorbox}[colframe=black, colback=white, coltitle=white, colbacktitle=black,
title=Prompt Template for Identifying Useful Code Spans,
boxrule=0.8pt,
fonttitle=\mdseries\small,
fontupper=\ttfamily\scriptsize,
enhanced,
rounded corners,
listing only,
breakable,
]
\refstepcounter{prompt}
\label{prompt3}
\begin{verbatim}
You are an autonomous AI assistant tasked with identifying relevant code in a codebase.  
Your goal is to select key code sections from the search results that are most relevant 
to the search request.

---

## Context

The previous messages will contain:

1. A search request from an AI assistant  
2. Search results containing various code sections with their line numbers  

---

## Your Task

### 1. Understand the Search Request
- Analyze the previous search request to understand what elements are being looked for  
- Identify key elements such as **functions, variables, classes, or patterns** that are 
relevant  

### 2. Evaluate Search Results
- Examine each code section in the search results for alignment with the search request  
- Assess the **relevance and importance** of each code section  
- Consider the **complete context** of code sections  

### 3. Respond with the Identify Action
- Select and respond with the **code sections** that best match the search request  
- Provide your **analysis in the thoughts field**  
- List the relevant **file paths** with **start and end line numbers** in the 
`identified_spans` field
\end{verbatim}
\end{tcolorbox}

\begin{tcolorbox}[colframe=black, colback=white, coltitle=white, colbacktitle=black,
title=Prompt Template for Node Evaluation,
boxrule=0.8pt,
fonttitle=\mdseries\small,
fontupper=\ttfamily\scriptsize,
enhanced,
rounded corners,
listing only,
breakable,
]
\refstepcounter{prompt}
\label{prompt4}
\begin{verbatim}
Your role is to evaluate the **last executed action** of the search tree that our AI
agents are traversing, to help determine the best trajectory to solve a programming 
issue. The agent is responsible for identifying and modifying the correct file(s) 
in response to the problem statement.

**Important:** While line numbers may be referenced in the initial problem description, 
they can shift as changes are made to the file. Focus on whether the agent is modifying 
the correct logical parts of the code, rather than strictly matching the initially 
mentioned line numbers. What matters is that the right section of code is being 
modified, even if its current line number differs from what was originally specified.

## Task

At this stage, the agent is still working on the solution. Your task is twofold:

1. **Evaluation**: Assess whether the change done by the **last executed action** is 
appropriate for addressing the problem and whether the agent is on the right path to 
resolving the issue. Verify that the correct sections of code are being modified, 
regardless of their current line numbers.  

2. **Alternative Feedback**: Independently of your evaluation, provide guidance for 
an alternative problem-solving branch. This ensures parallel exploration of different 
solution paths.

## Evaluation Criteria

- **Exploratory Actions:** Initial searches and information-gathering steps are 
essential and should not be heavily penalized if they don't yield immediate results.  
- **Appropriateness of Action:** Is the action logical given the agent's current 
knowledge and early stage of problem-solving?  
- **Query Relevance:** Is the search query or parameters well-defined and likely to 
find relevant code?  
- **Search Scope Appropriateness:** Do the file patterns and class/function names 
narrow down the search effectively?  
- **Relevance of Search Results:** Are the search results directly related to the 
problem and useful for progress?  
- **Size of Search Results:** Is the code context appropriately sized (not too large 
or too small)?  
- **Category Appropriateness:** Does the category (implementation or test) align with 
the search intent?

## Reward Scale

Assign an integer between 0 and 100:

- 90-100: Excellent action; parameters are well-defined and results are highly 
relevant.  
- 75-89: Good action; parameters are reasonable and results are relevant.  
- 25-74: Action has issues or yields few/no relevant results.  
- 0-24: Counterproductive; results are irrelevant or excessively large.

## Feedback Structure

- **Explanation:** Provide detailed reasoning for your decision, focusing on the **last 
executed action**, its relation to previous actions, and its impact.  
- **Feedback to Alternative Branch:** Suggest conceptual alternative approaches without 
actual code, avoiding actions that would replicate previous outcomes.  
- **Reward:** Assign a single integer between 0 and 100 based on confidence in 
correctness and likelihood of solving the issue.

## Available Actions

### FindClass
Use this when you know the exact class name to find.

- Finding class implementations: `class_name="UserRepository"`  
- Locating test classes: `class_name="TestUserAuthentication"`  
- Finding base classes: `class_name="BaseController"`  
- Classes in specific modules: `class_name="Config", file_pattern="src/config/*.py"`

### FindFunction
Use when you know the exact function or method name.

- Finding test cases: `function_name="test_user_login"`  
- Locating implementations: `function_name="process_payment"`  
- All methods with a name: `function_name="validate"`  
- Specific class method: `function_name="save", class_name="UserRepository"`

### FindCodeSnippet
Use when you know the exact code snippet.

- Finds constants: `code_snippet="MAX_RETRIES = 3"`  
- Finds decorators: `code_snippet="@retry(max_attempts=3)"`  
- Finds imports: `code_snippet="from datetime import datetime"`  
- Configuration patterns: `code_snippet="DEBUG = os.getenv('DEBUG', False)"`

> Note: If you don't know the exact code, use **SemanticSearch**.

### SemanticSearch
Use when you don't know exact names but want related functionality.

- Functionality by description: `query="code that handles password hashing"`  
- Related test cases: `query="tests for user registration", category="test"`  
- Implementations: `query="database connection pooling", category="implementation"`  
- Error handling patterns: `query="error handling for API requests"`

> Flexible for unknown names, discovering related code, or exploring features.

### ViewCode
View the code in a file or a specific span.

### Finish
Indicate that the generated code answer is accurate and complete for the user's query.
\end{verbatim}
\end{tcolorbox}

\begin{tcolorbox}[colframe=black, colback=white, coltitle=white, colbacktitle=black,
title=Prompt for LLM-as-Judge,
boxrule=0.8pt,
fonttitle=\mdseries\small,
fontupper=\ttfamily\scriptsize\linespread{1.12}\selectfont,
enhanced,
rounded corners,
listing only,
breakable,
]
\refstepcounter{prompt}
\label{prompt5}
\begin{verbatim}
You are a STRICT and RIGOROUS evaluator. You must rate the candidate answer STRICTLY against the reference answer. 
Be CONSERVATIVE with high scores - only award high scores (16-20) when the candidate answer is truly excellent and 
closely matches the reference answer in quality and content.

CRITICAL EVALUATION PRINCIPLES:
1. Compare the candidate answer DIRECTLY with the reference answer point by point
2. Any deviation, omission, or inaccuracy should result in score reduction
3. High scores (16-20) should be RARE - reserve them only for answers that are nearly perfect
4. Be strict about factual accuracy - even minor errors should lower the correctness score
5. Missing key points from the reference answer should significantly reduce completeness score
6. Vague or imprecise language should lower clarity scores
7. When in doubt between two score ranges, choose the LOWER one

Evaluation Criteria and Scoring Guidelines (each scored 1 to 20, total score 100):
    1. Correctness (STRICT - penalize any inaccuracies):
        20 --- ONLY if completely correct with ALL core points and details accurate, matching reference answer 
        precisely
        16-19 --- Mostly correct but must have only TRIVIAL inaccuracies; any noticeable error reduces to 15 or below
        12-15 --- Partially correct; has some errors or omissions that affect understanding; main points may be 
        accurate but details are wrong
        8-11 --- Several errors or ambiguities that significantly affect understanding of core information
        4-7 --- Many errors; misleading or fails to convey key information correctly
            1-3 --- Serious errors; completely wrong or misleading
    2. Completeness (STRICT - penalize missing information):
        20 --- ONLY if covers ALL key points from reference answer without ANY omission; must match reference in depth
        16-19 --- Covers most key points but missing some non-trivial information; minor omissions are acceptable
        12-15 --- Missing several important key points; content is noticeably incomplete compared to reference
        8-11 --- Important information largely missing; content is one-sided or superficial
        4-7 --- Covers very little relevant information; seriously incomplete
        1-3 --- Covers almost no relevant information; completely incomplete
    3. Relevance (STRICT - penalize off-topic content):
        20 --- ONLY if content is fully focused on question topic with NO irrelevant information whatsoever
        16-19 --- Mostly focused but may have minor peripheral information; any significant off-topic content reduces 
        score
        12-15 --- Generally on topic but contains some off-topic content that detracts from answer
        8-11 --- Topic not sufficiently focused; contains considerable off-topic or tangential content
        4-7 --- Content deviates from topic; includes excessive irrelevant information
        1-3 --- Majority of content irrelevant to the question
    4. Clarity (STRICT - penalize unclear expression):
        20 --- ONLY if language is exceptionally fluent, clear, and precise; very easy to understand without any 
        ambiguity
        16-19 --- Mostly fluent and clear but may have minor unclear points; any significant ambiguity reduces score
        12-15 --- Generally clear but some expressions are unclear or not concise; may require effort to understand
        8-11 --- Expression somewhat awkward; has ambiguity or lacks fluency that hinders understanding
        4-7 --- Language obscure; sentences are not smooth; significantly hinders understanding
        1-3 --- Expression confusing; very difficult to understand
    5. Reasoning (STRICT - penalize weak logic):
        20 --- ONLY if reasoning is exceptionally clear, logical, and well-structured; argumentation is excellent and 
        matches reference quality
        16-19 --- Reasoning is clear and logical with solid argumentation; minor logical gaps may exist
        12-15 --- Reasoning generally reasonable but has noticeable logical jumps or organization issues
        8-11 --- Reasoning is average; has logical jumps or organization problems that affect understanding
        4-7 --- Reasoning unclear; lacks logical order; difficult to follow
        1-3 --- No clear reasoning; logic is chaotic

INPUT:
    Question:{question}
    Reference Answer:{reference}
    Candidate Answer:{candidate}

OUTPUT:
    Please output ONLY a JSON object with 5 integer fields in the range [1,20], corresponding
    to the evaluation scores:
        {{
        "correctness": <1-20>,
        "completeness": <1-20>,
        "relevance": <1-20>,
        "clarity": <1-20>,
        "reasoning": <1-20>
        }}

SCORING INSTRUCTIONS:
- Read the reference answer carefully and identify ALL key points, details, and structure
- Compare the candidate answer systematically against the reference answer
- For each criterion, start with a conservative score and only increase if the candidate truly deserves it
- If the candidate answer is significantly shorter, less detailed, or less precise than the reference, reduce 
scores accordingly
- If the candidate answer contains information not in the reference (unless it's clearly relevant and accurate), 
consider reducing relevance score
- When scoring, ask yourself: "Does this can
didate answer match the quality and completeness of the reference 
answer?" If not, reduce scores
- Average or mediocre answers should receive scores in the 8-15 range, not higher
- Only truly excellent answers that closely match the reference should receive 16-20 scores

REQUIREMENT:
    No explanation, no extra text, no formatting other than valid JSON. Be strict and conservative with your scores.
\end{verbatim}
\end{tcolorbox}

\ifdefined\DeepRepoQASupplementInput
\else
\end{document}
\fi

\end{document}